\documentclass[traditabstract]{aa}
\usepackage{graphicx}
\usepackage{txfonts}
\usepackage{todonotes}
\usepackage[normalem]{ulem}
\usepackage{amstext}
\usepackage[breaklinks=true]{hyperref}
\usepackage{xspace}
\usepackage{lscape}
\usepackage{mathrsfs}
\usepackage{color}
\usepackage{placeins}

\usepackage{listings}
\usepackage[none]{hyphenat}

\makeatletter
\renewcommand*\maketitle{%
  \thispagestyle{firstpage}
\begingroup
    \if@wideboxfn
    \setlength\bibindent{1.4\parindent}
    \else
    \setlength\bibindent{\parindent}
    \fi
    \renewcommand*\thefootnote{\@fnsymbol\c@footnote}%
    \renewcommand\@makefntext[1]{%
    \ifaa@longfn\hsize\textwidth\fi
    \noindent
    \hb@xt@\bibindent{\hss\@makefnmark\enspace}##1}
  \ifaa@twocolumn
  \begingroup
    \begin{aa@strip}
          \aa@maketitle
    \end{aa@strip}
    \@thanks            
  \endgroup
  \else
    \begingroup
      \let\thanks\footnote
      \aa@maketitle
    \endgroup
  \fi
\endgroup
  \setcounter{footnote}{0}%
}
\makeatother

\definecolor{dkgreen}{rgb}{0,0.6,0}
\definecolor{gray}{rgb}{0.5,0.5,0.5}
\definecolor{mauve}{rgb}{0.58,0,0.82}
\definecolor{red}{rgb}{1,0,0}
\newcommand{\orcit}[1]{\protect\href{https://orcid.org/#1}{\protect\includegraphics[width=8pt]{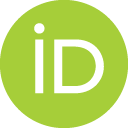}}}

\def\ltsima{$\, \buildrel < \over \sim \,$}
\def\simlt{\lower.5ex\hbox{\ltsima}}
\def\gtsima{$\, \buildrel > \over \sim \,$}
\def\simgt{\lower.5ex\hbox{\gtsima}}

\begin{document}

\title{RGB Tip distance to the faint gas-rich  dwarf KK~153
} 
\authorrunning{M. Bellazzini et al.}
\titlerunning{RGB Tip distance to KK~153}

\author{
M.~Bellazzini\orcit{0000-0001-8200-810X}\inst{1}
\and
G.~Beccari\orcit{0000-0002-3865-9906}\inst{2}
\and
R.~Pascale\orcit{0000-0002-6389-6268}\inst{1}
\and
D.~Paris\orcit{0000-0002-7409-8114}\inst{3}
\and
F.~Annibali\orcit{0000-0003-3758-4516}\inst{1}
\and
F.~Cusano\orcit{}\inst{1}
\and
D.~P\'erez-Mill\'an\orcit{0000-0002-4507-9571}\inst{1}
}
\institute{
INAF - Osservatorio di Astrofisica e Scienza dello Spazio di Bologna, via Piero Gobetti 93/3, 40129 Bologna, Italy
\and
European Southern Observatory, Karl-Schwarzschild-Strasse 2, 85748 Garching bei M\"unchen, Germany
\and
INAF - Osservatorio Astronomico di Roma, Via Frascati 33, 00078, Monteporzio Catone, Rome, Italy
}

\date{Accepted for publication by A\&A}

\abstract{KK~153 is a star-forming dwarf galaxy that has been recently proposed as a new member of the sparsely populated class of gas-rich ultra faint dwarfs, lying in the outskirts of the Local Group. We used the Large Binocular Telescope under sub-arcsec seeing conditions to resolve for the first time the outer regions of KK~153 into individual stars, reaching the red giant branch. The magnitude of the red giant branch tip was used to measure a distance of $D=3.06^{+0.17}_{-0.14}$~Mpc, much more accurate and precise than the estimate previously available in the literature, based on the baryonic Tully-Fisher relation ($D=2.0^{+1.7}_{-0.8}$~Mpc). The new distance places KK~153 clearly beyond the boundaries of the Local Group, and, together with a new measure of the integrated magnitude, implies a stellar mass of ${\rm M_{\star}}=2.4 \pm 0.2 \times 10^6~{\rm M_{\sun}}$. The dwarf populates the extreme low-mass tail of the $M_{\star}$ distribution of gas-rich galaxies but it is significantly more massive than the faintest local gas-rich dwarfs, Leo~T and Leo~P. In analogy with similar systems, the star formation history of KK~153 may have been impacted by the re-ionisation of the Universe while keeping a sufficient gas reservoir to form new stars several Gyr later.
}


\maketitle

\section{Introduction}
\label{sec:intro}

The exploration of the very faint end of the galaxy luminosity function is a fascinating endeavour that can be pursued only in the local Universe, where extremely faint stellar systems can be actually detected \citep{belo13,simon19}. One of the techniques that has been used in the last decade to look for very faint nearby galaxies, down to the regime of {\it almost dark} \citep[][]{cannon15,giova15,sand15, leisman17,leisman21,jones23} or, possibly, completely dark systems \citep{anand25}, has been to search for stellar counterparts of compact \ion{H}{I} clouds whose low line-of-sight (los) velocity is compatible with a local nature \citep[see, e.g.,][for recent catalogues of such clouds]{adams13,saul14}. Systematic searches led to the discovery of a handful of gas-rich low-mass stellar systems \citep[see, e.g.,][]{cannon15,secco,mic15,sand15}, the most intriguing case being probably Leo~P, a low-mass very metal-poor star-forming dwarf galaxy at $D\simeq 1.6$~Mpc, likely associated with the NGC~3109 group \citep{giova13,skill13,mcquinn15,mcquinn24}.

Along this line of research, \citet{xu25} recently identified the stellar counterpart of a low-velocity compact \ion{H}{i} cloud discovered in the Five-hundred-meter Aperture Spherical radio Telescope (FAST) extragalactic \ion{H}{i} survey \citep{zhang24}. In spatial coincidence with the \ion{H}{i} cloud they found a small bluish galaxy of elongated shape, KK~153 (LEDA~4192). \citet{xu25}, having carefully analysed all the available data concluded that KK~153 has a stellar mass near the threshold adopted to divide ordinary dwarf galaxies and Ultra Faint Galaxies \citep[UFDs; $M_{\star}\simeq 10^5~M_{\sun}$;][]{mcquinn24}, thus suggesting that it may be a new member of the rare class of gas-rich UFDs in the Local Group or its immediate vicinity, together with  Leo~T \citep[$D=0.4$~Mpc][]{irwin07} and Leo~P \citep[$D=1.6$~Mpc,][]{mcquinn15}.

However, many of the conclusions drawn by \citet{xu25} were based on a highly uncertain distance estimate, based on the baryonic Tully-Fisher relation, $D=2.0^{+1.7}_{-0.8}$. In our experience, a much more accurate distance measure for dwarfs in this distance range can be very efficiently obtained with less than half an hour of observation with the Large Binocular Camera \citep[LBC, mounted at the Large Binocular Telescope\footnote{LBT; \url{https://www.lbto.org}};][]{lbc} in binocular mode, under sub-arcsec seeing conditions \citep[][]{vv124, secco,ssh_pap1,sacchi24}, by resolving the outskirts of these galaxies into individual stars to a depth where the tip of the Red Giant Branch (RGB) can be used as a standard candle \citep[see, e.g.,][and references therein]{lee93, salacas98, madore20, bptip24}. For this reason we asked to observe KK~153 with LBC under the Italian LBT Director Discretionary Time and we were awarded time under the programme DDT-2024B-03. 

In this paper we describe the results of this experiment, that lead to a significantly revised and improved measure of the distance to KK~153. Albeit our new measure locates KK~153 clearly beyond the Local Group, implying a larger stellar mass than that inferred by \citet{xu25}, KK~153 remains among the least massive local and isolated gas-rich dwarfs, sampling the vicinity of the mass threshold of systems that were able to survive quenching by the background UV radiation field that re-ionised the Universe, a crucial regime to understand the evolution of dwarf galaxies \citep[see, e.g., and references therein][]{gutcke22,xu25,pavo_dist,mcquinn24,jones25}.

The paper is organised as follows: in Sect.~\ref{sec:obs} we describe the observations, the data reduction and calibration, the adopted selection and the Colour Magnitude Diagram (CMD) of KK~153 that we obtained from our data; in Sect.~\ref{sec:dist} we derive a new distance measure to KK~153 using the RGB Tip as a standard candle and we provide new measures of the integrated magnitudes and estimates of the stellar mass. Finally, in Sect.~\ref{sec:discu} we summarise and briefly discuss our results. The most relevant properties of KK~153 derived in this paper are summarised in Table~\ref{tab:tab1}.

\begin{table}[!htbp]
\centering
\caption{\label{tab:tab1} Main properties of KK~153.}
{
    \begin{tabular}{lll}
Parameter  & Value & Notes \\
\hline 
$(m-M)_0$& ${ 27.43^{+0.12}_{-0.10}}$ & this work \\
$D$      & ${ 3.06^{+0.17}_{-0.14}}$~Mpc & this work \\
$E(B-V)$ & $0.009\pm 0.001$ & SFD98+SF10\\
r$_0$  &$16.35\pm0.14$ & integrated, t.w. \\
g$_0$  &$16.86\pm0.09$ & integrated, t.w. \\
$R_h$      & $17.4\arcsec\pm0.8\arcsec$  & (major axis) this work     \\
$M_V$  & $-10.87\pm0.20$ & this work  \\
$M_{\star}$  & $2.4\pm 0.2\times 10^6~M_{\sun}$  & this work  \\
M$_{\ion{H}{I}}$  & $1.2\pm 0.4 \times10^6~M_{\sun}$  & t.w. + \citet{xu25}  \\
$ell=(1-b/a)$  &     0.4              & HyperLeda$^a$ \\
PA     &$167.0\degr \pm 1.7\degr$ & \citet{xu25} \\
$V_h$  &$127.5\pm0.2$ km~s$^{-1}$ & \citet{xu25} \\
\hline
    \end{tabular}
}
\tablefoot{$^a$ \url{http://atlas.obs-hp.fr/hyperleda/}.
$V_h$ is the heliocentric 
line of sight velocity, measured with \ion{H}{i} observations.
}
\end{table}

\section{Observations and data reduction}
\label{sec:obs}
\begin{figure}[ht!]
\center{
\includegraphics[width=\columnwidth]{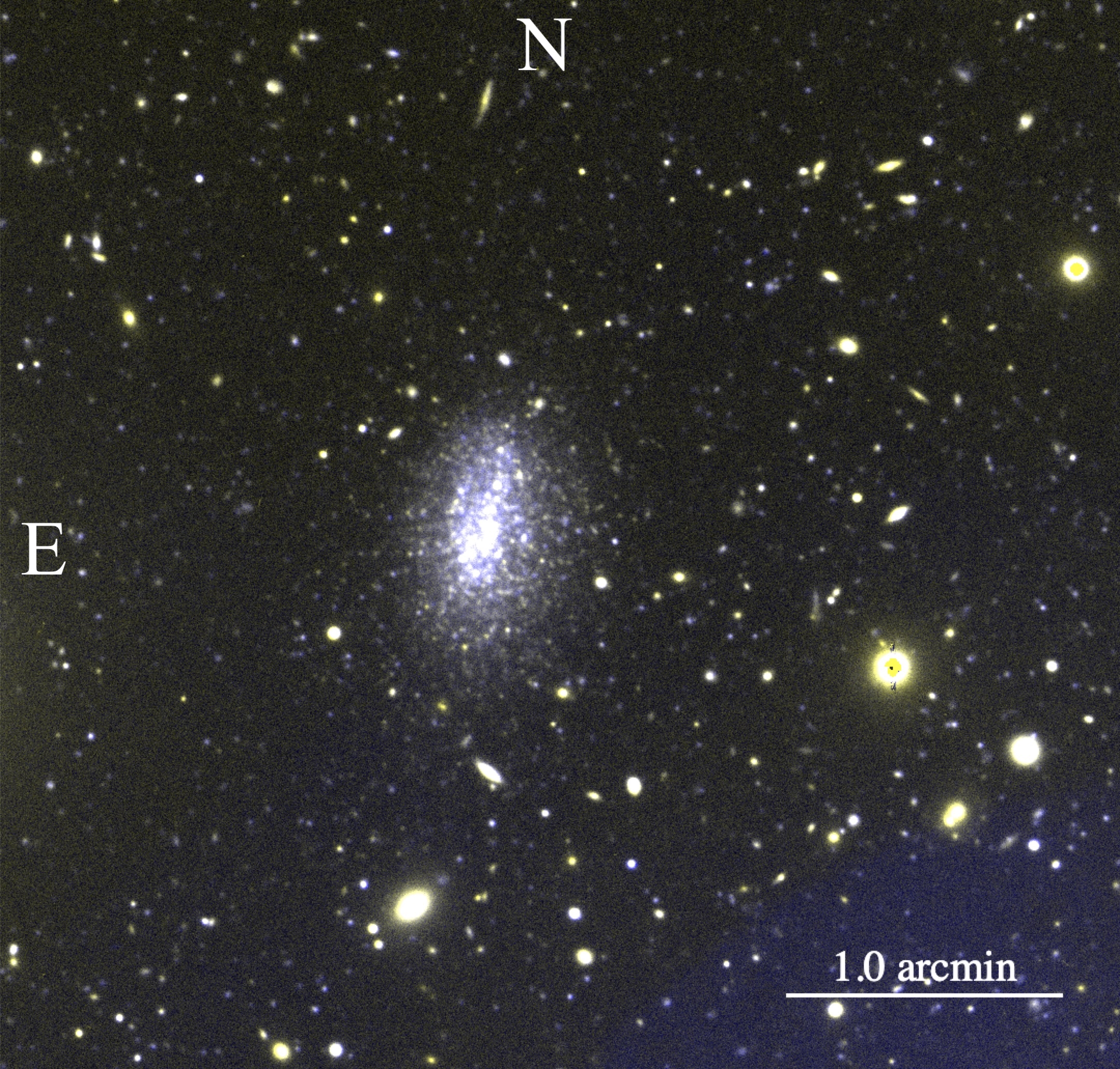}
}
\caption{Red-Green-Blue cut-out image of KK~153 and its immediate surroundings obtained from sky-subtracted, stacked deep g and r images. The r image has been used for both the Red and the Green channels, while the g image has been used for the Blue channel.} 
\label{fig:ima}
\end{figure}

Observations were carried out with the LBC in binocular mode, providing simultaneous imaging in the Sloan Digital Sky Survey\footnote{\url{https://www.sdss.org}} (SDSS) $g$ and $r$ bands. The data were acquired on May 21, 2025 (UT) under clear sky conditions, with an average seeing of $0.9^{\prime\prime}$. The total exposure time per filter was 1800\,s, split into six individual 300\,s exposures. The images were bias-subtracted, corrected for flat-field and cleaned from hot and dead pixels as well as for cosmic ray hits with the standard procedure adopted by the INAF LBT-Italia team\footnote{\url{https://lbt.inaf.it}} \citep[see, e.g.,][for a detailed description]{ssh_pap1}. A portion of an image obtained from the stacking of all the $g$ and $r$ images centred on KK~153 is shown in Fig.~\ref{fig:ima}. A population of bright blue stars broadly clustered in the central regions, along the minor axis, is surrounded by an elongated and more extended  halo of fainter and redder stars, fully consistent with the findings by \citet{xu25}. 

LBC is a mosaic of four adjacent 4608~px$\times$2048~px CCDs, with a pixel scale of 0.225$\arcsec{\rm px^{-1}}$ \citep{lbc}. Each CCD chip covers a field of $17.3\arcmin\times 7.7\arcmin$. In our images, KK~153 is imaged near the center of the central chip (Chip~2; slightly offset to minimise the contaminating effect of a few very bright stars in the field). Since the field of view of Chip~2 is wide enough to sample KK~153 and an ample portion of the surrounding field, in the following we limit our analysis to Chip~2 images, as done in \citet{vv124} in the study of a similar system.

We performed Point Spread Function (PSF) photometry of individual stars using DAOPHOT~II~\citep[][]{stet87}. { In summary, 20 to 30 well-sampled, not saturated and isolated stars were selected in each individual frame to estimate the PSF models.} The best PSF models were { fit} using a Lorentz function. The spatial trends of the PSF shape within the frame were fitted with a second order polynomial in x,y. The  PSF fitting was performed on each image using the ALLFRAME routine~\citep[][]{allframe}. We first created a master catalogue of sources selecting the objects with peaks higher than $3\sigma$ above the background in a stacked image obtained by registering and co-adding all the images considered for the analysis. Then, ALLFRAMES automatically identifies and fits each of these sources on the individual images. Only sources found at least in three $g$ and three $r$ images were retained in the final catalogue. The average and the standard error of the mean of the independent measures obtained from the different images were adopted as the final values of the instrumental magnitude and of the uncertainty on the relative photometry. 

\begin{figure}[ht!]
\center{
\includegraphics[width=\columnwidth]{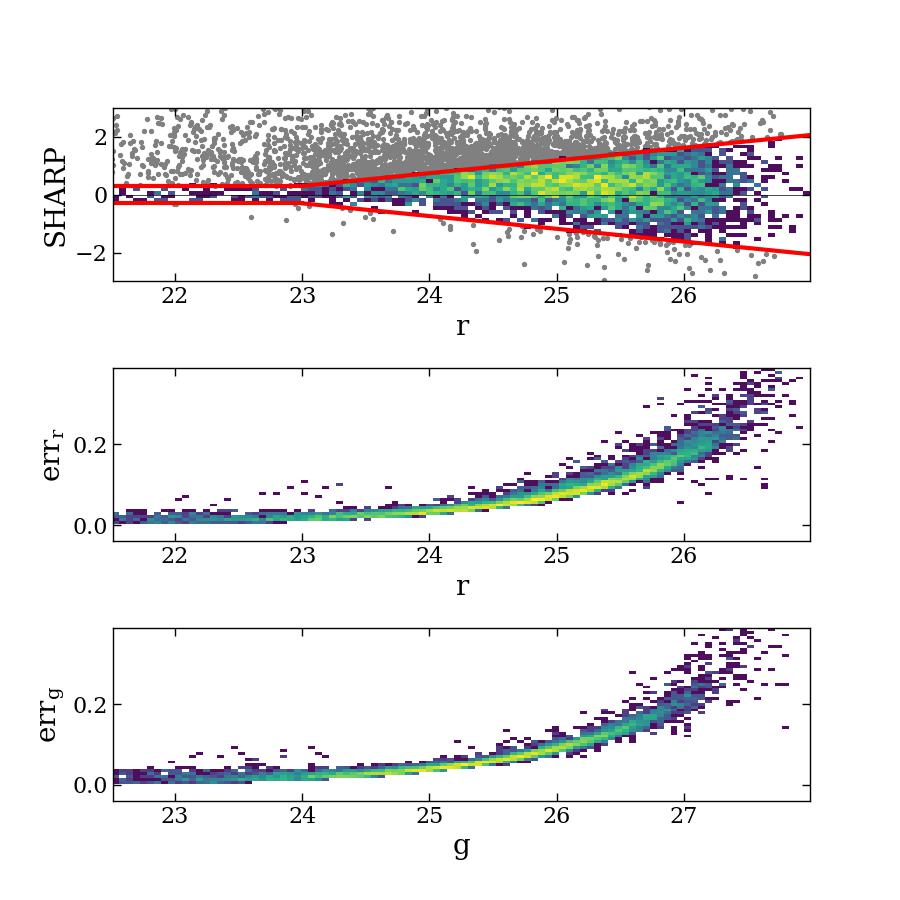}
}
\caption{Upper panel: distribution of the Daophot SHARP shape parameter as a function of $r$ magnitude. { Sources lying in the region enclosed by the red contours are considered as likely stars, while those outside the contours (grey points), mostly distant unresolved galaxies, have been excluded from the subsequent analysis}. Middle and lower panels: distribution of the photometric errors in $r$ and $g$ as a function of $r$ and $g$ magnitude, respectively. } 
\label{fig:sel}
\end{figure}

In wide-field deep stellar photometry, distant background galaxies are the major source of contamination \citep[see][for a detailed discussion in a very similar context and for further references]{vv124}. To mitigate the contamination of our sample by these sources and by other spurious detections we adopted the magnitude-dependent selection on the DAOPHOT~II $SHARP$ parameter illustrated in the upper panel of Fig.~\ref{fig:sel}\footnote{ In particular we retain all the sources having $r<=23.0$ and $|SHARP|<0.3$, or $r>23.0$ and $|SHARP|<0.44r-9.82$. The selection window has been drawn by eye to reject most of the obvious outlier of the distribution of SHARP as a function of magnitude, with particular care for the rejection of marginally extended objects.}. $SHARP$ should take values as near to zero as the shape of the light distribution is more similar to that expected from a point-like source (PSF), while significant deviations toward positive values indicates extended sources (e.g, galaxies or blended stars) and deviations toward negative values identify sources more compact than point sources (like, e.g., residual small cosmic spikes or hot pixels). The adopted selection should remove a large fraction of contaminating background galaxies, with efficiency decreasing toward fainter magnitudes, as expected.

\begin{figure}[ht!]
\center{
\includegraphics[width=\columnwidth]{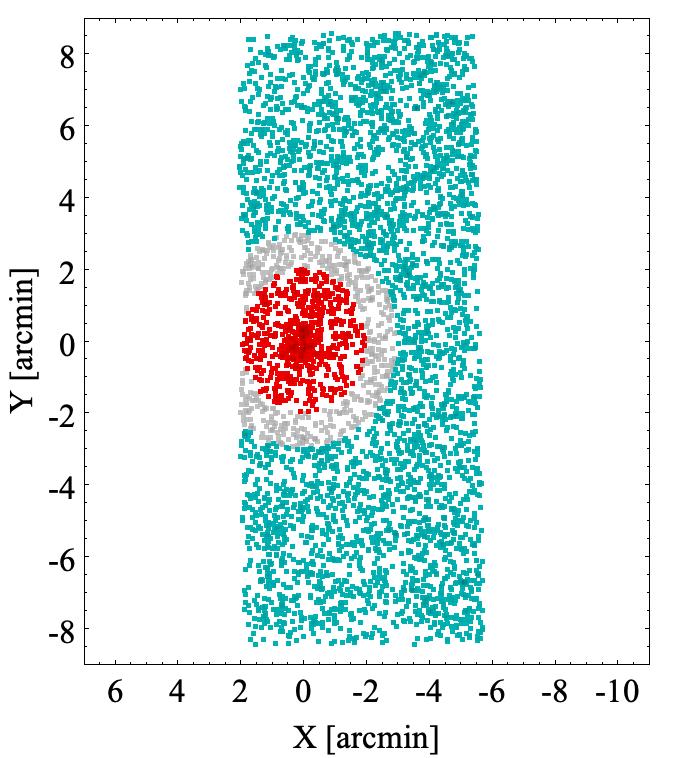}
}
\caption{Sky map of the position of the sources in our selected catalogue, in an orthographic projection reference system with the origin at the center of KK~153. North is up, East to the left. Sources within $2.0\arcmin$ from the centre of KK~153 are plotted in red (galaxy), sources with $R>3.0\arcmin$ are plotted in cyan (field).} 
\label{fig:map}
\end{figure}

The middle and lower panels of Fig.~\ref{fig:sel} show the distribution of photometric errors as a function of the magnitude, in $r$ and $g$. The photometric error is $\le 0.05$~mag for $g(r)\la 25.0(24.0)$, then it gently increases, reaching $\simeq 0.2$~mag at $g(r)\simeq 27.0(26.0)$. To further clean the final catalogue on which we performed the scientific analysis, in addiction to the selection in $SHARP$ described above, we rejected all the sources having $err_g\ge 0.2$ or $err_r\ge 0.2$. The distribution in the sky of the finally selected sources is presented in Fig.~\ref{fig:map}, where we plot in red the sources within $R=2.0\arcmin$ from the centre of KK~153 \citep[taken from][]{xu25} and in cyan those farther that $R=3.0\arcmin$, as, in the following, these two regions will often serve as reference for the KK~153 "on target" sample and for the Control Field, respectively. The over-density of sources corresponding to the resolved stars in KK~153 is readily evident at the center of the red circular region. 

\subsection{Photometric and astrometric calibrations}

The astrometric and photometric calibrations have been achieved using sources in common with the SDSS-DR16 dataset \citep{ahumada20}. The astrometry of our raw catalogue has been mapped into the SDSS-DR16 system with a 4th order polynomial fitted on 112 stars in common, with the appropriate code of CataPack\footnote{\url{http://davide2.bo.astro.it/~paolo/Main/CataPack.html}} suite. The resulting rms spread is $<0.15\arcsec$ both in RA and DEC.

The photometric calibration has been obtained from the subset of the brightest sources classified as stars in SDSS-DR16, about 50 stars covering the colour range enclosing the large majority of the stars of scientific interest, $0.0\la g-r\la 1.8$. A first order polynomial as a function instrumental g-r colour has been adopted, strictly analogous to what done in \citet{vv124}, \citet{secco}, and \citet{ssh_pap1}. In the following, the rms spread about the calibrating relations, 0.073~mag and 0.076~mag in g and r band, respectively, will be added in quadrature to the error budget of all the derived quantities that depend on the absolute photometric calibration.

\subsection{Surface photometry}
\label{sec:apphot}

We performed the integrated photometry of the galaxy using the Aperture Photometry Tool \citep[APT;][]{apt1,apt2} on our stacked $g$ and $r$ images. Having verified that the axis ratio reported in the Hyperleda\footnote{\url{http://atlas.obs-hp.fr/hyperleda/}} catalogue \citep{hyperleda} and the position angle reported by \citet{xu25} are appropriate for KK~153 (see Table~\ref{tab:tab1}), we adopted elliptical apertures accordingly shaped and oriented. 

To measure the integrated photometry we adopted an aperture with semi-major axis of $180$~px$=40.5\arcsec$, that should enclose most of the KK~153 light. The sky background was computed on a concentric elliptical annulus with inner(outer) semi-major axis of 300~px(350~px), width=11.25$\arcsec$, wide enough to account for the contribution of background/foreground sources to the surface photometry over such large apertures. 

Our integrated magnitudes are $\simeq 0.5$~mag brighter than those reported by \citet[][]{xu25}. No detail on the way in which their integrated magnitudes have been obtained is reported in \citet{xu25}. A possible explanation for the observed difference is that they may have adopted a smaller aperture. With the same aperture used above, we obtain results fully compatible with those from LBC images, within the uncertainties, from photometrically calibrated SDSS images, thus supporting our measures. The rationale behind the
choice of a large aperture, was driven by the fact that we detect RGB stars attributable to KK~153 at least out to R$\simeq 1.5\arcmin$ from the centre of the galaxy (see below). The finally adopted aperture size is trade-off between the extension of the stellar distribution and the requirement of avoiding the inclusion of an excessively extended region where the light from the galaxy is below the sky level.

Finally, we empirically estimated the half-light radius as the semi-major axis of the elliptical aperture containing half of the flux measured in the aperture used for the integrated magnitude. The value reported in Table~\ref{tab:tab1} is the average of the values we got from the $g$ and $r$ images, the uncertainty is the associated standard deviation.

\begin{figure*}[ht!]
\center{
\includegraphics[width=0.9\textwidth]{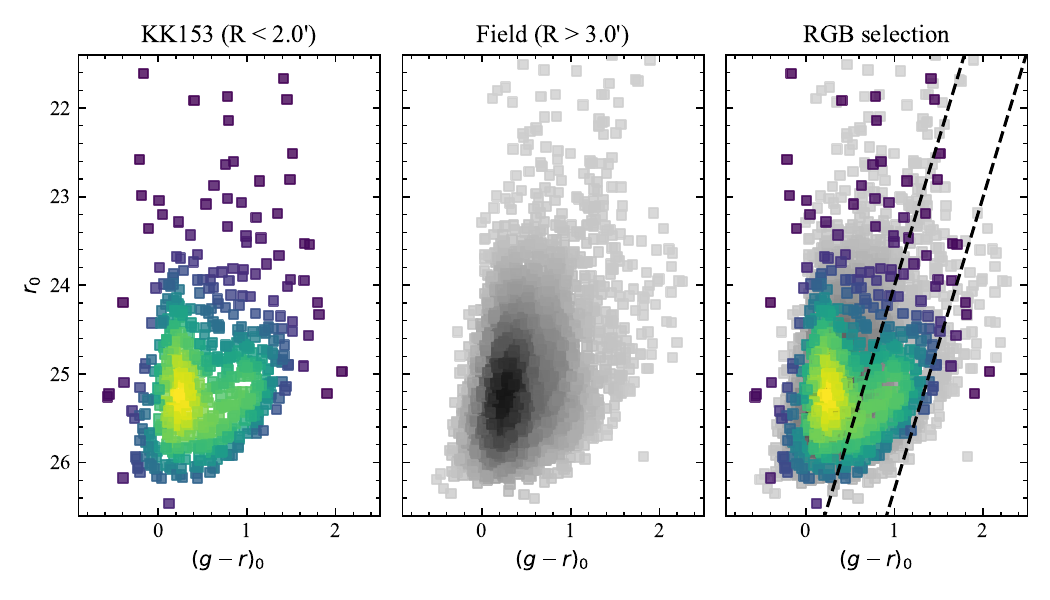}
}
\caption{Left panel: CMD of stars enclosed within a circle of radius=2.0$\arcmin$ centred on the center of KK~153. Middle panel: CMD of all the stars in our field located at an angular distance larger than =3.0$\arcmin$ from the centre of KK~153 (Field). It is worth noting that the area of the $R>3.0\arcmin$ field is more than 8.3 times the area of the $R<2.0\arcmin$ circle around the galaxy. Right panel: the CMD of KK~153 superimposed to the CMD of the Field population. The two dashed parallel lines display our selection of candidate RGB stars of KK~153. In all the CMDs the colour intensity of the points is proportional to the local star count density (Hess diagram).}
\label{fig:cmd}
\end{figure*}

\subsection{The Colour Magnitude Diagram}
\label{sec:cmd}

The CMDs of the sources enclosed within $2.0\arcmin$ (galaxy) and of those lying farther than $3.0\arcmin$ from the centre of KK~153 (field) are shown in the left and middle panels of Fig.~\ref{fig:cmd}, respectively. The "field" stars are distributed over a sky area that is more than 8.3 times the area covered by the "galaxy" sample. All the magnitudes are corrected for foreground extinction adopting $A_g=3.734E(B-V)$ and $A_r=2.583E(B-V)$, following \citet{bptip24}, and $E(B-V)=0.009$ from the \citet[][SFD98]{sfd98} reddening maps recalibrated according to \citet[][SF10]{schlafly2011}. We used the same maps to verify that the extinction variations are negligible across the considered FoV. 

For $r_0\ge 23.5$, the CMD of the field is dominated by the wedge-shaped distribution of distant galaxies in the blue sequence, around $(g-r)_0\simeq 0.2$, while most of the sources redder than $(g-r)_0\simeq 1.0$, at any magnitude, are likely local M dwarfs, and most bluer sources at $r_0\le 23.5$ are Galactic main sequence (MS) stars along the line of sight \citep[][]{vv124}. In the $R<2.0\arcsec$ CMD the blue sequence wedge is still the dominant feature, but a handful of stars bluer than $(g-r)_0\simeq 0.0$ also emerges, having no counterpart in the field: these are young stars located in the innermost corona of KK~153 that our observations are able to resolve into stars. An over-density of stars in the colour range $0.8\le (g-r)_0\le 1.5$ is also apparent, when compared to the "field" CMD. In analogy to the cases shown and discussed by \citet{sacchi24} in a very similar context, we identify this wide sequence as the upper RGB of KK~153, tipping about $r_0\simeq 24.5$. In the right panel of Fig.~\ref{fig:cmd} we illustrate how we selected candidate RGB stars of KK~153 in the following analysis.

\begin{figure}
\center{
\includegraphics[width=0.95\columnwidth]{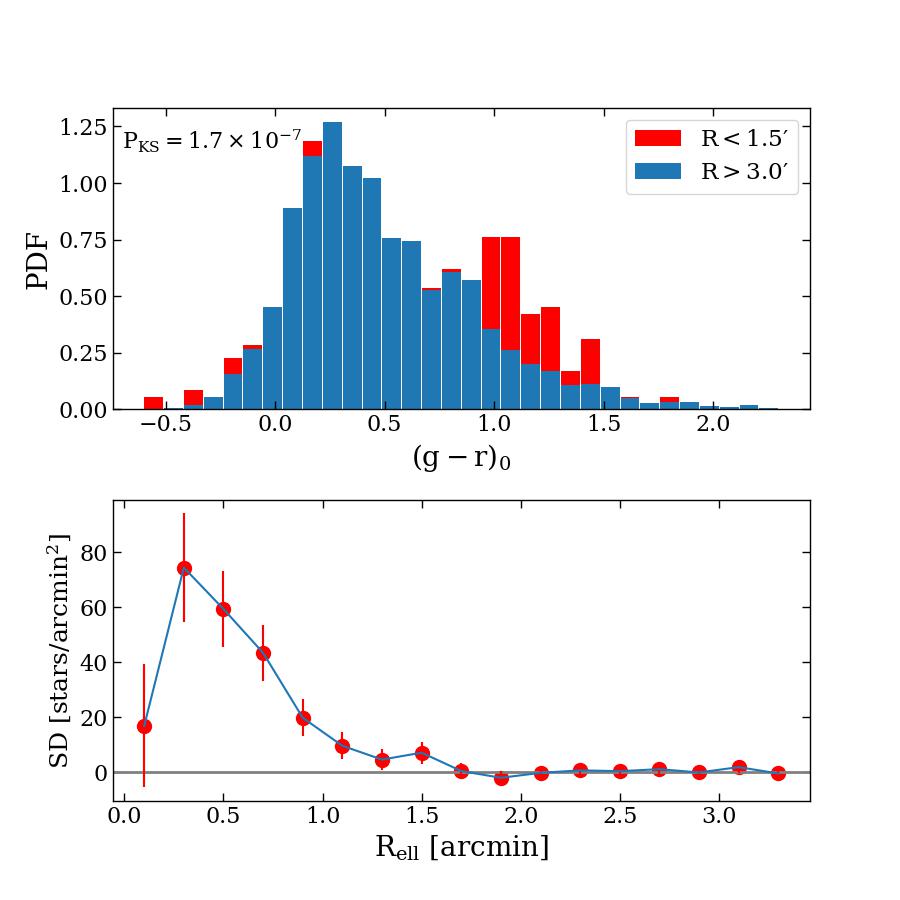}
}
\caption{Upper panel: the colour distribution of the stars within 1.5$\arcmin$ from the centre of KK~153 (red histogram) is compared to the colour distribution of the Field stars ($R>3.0$\arcmin; blue histogram). Lower panel: background-subtracted surface density profile of RGB stars (selected as in Fig.~\ref{fig:cmd}, with the additional condition $r_0>23.0$) from the centre of KK~153.}
\label{fig:colprof}
\end{figure}

However, given the relatively low number of likely KK~153 stars that we actually resolve, before proceeding with the measure of the distance, we want to provide additional evidence that we have indeed positively detected the RGB of the galaxy. The comparison between the colour distribution of the "galaxy", here defined as the $R<1.5\arcmin$ circle to maximise the contrast, and "field" samples in upper panel of Fig.~\ref{fig:colprof}, normalised to their areas, clearly shows that in the galaxy there is a strong excess of stars in the range $1.0\le (g-r)_0\le 1.5$ with respect to the field. Limiting the comparison to stars fainter than $r_0=23.0$, we found that, according to the Kolmogorov-Smirnov test, the probability that the two distributions are drawn from the same parent population is as low as $P_{KS}=1.7\times 10^{-7}$.

Then, selecting the candidate RGB stars as shown in the left panel of Fig.~\ref{fig:cmd}, with the additional condition $r_0>23.0$, we derive the surface density profile of these stars about the centre of KK~153, along the major axis, using the elliptical radius $R_{ell}$ as defined in Eq.~1 of \citet{perina09}. The profile, shown in the lower panel of Fig.~\ref{fig:colprof}, demonstrate that the stars that we identify as RGB members of KK~153 are highly over-dense around the centre of the galaxy and are distributed along a coherently declining profile, over a scale of 4-5 half light radii\footnote{The drop of the surface density in the innermost point is due to the strong incompleteness affecting the central, overcrowded regions of the galaxy in our images.}. The difference in the surface density of RGB stars between the $R<1.5(2.0)\arcmin$ and the $R>3.0\arcmin$ regions exceeds zero by more than $8.6\sigma(8.3\sigma)$.
Hence, there is no doubt that we have resolved the brightest $\simeq 2$~mag of the RGB of KK~153.

\section{Distance and stellar mass}
\label{sec:dist}

\begin{figure}[ht]
\center{
\includegraphics[width=\columnwidth]{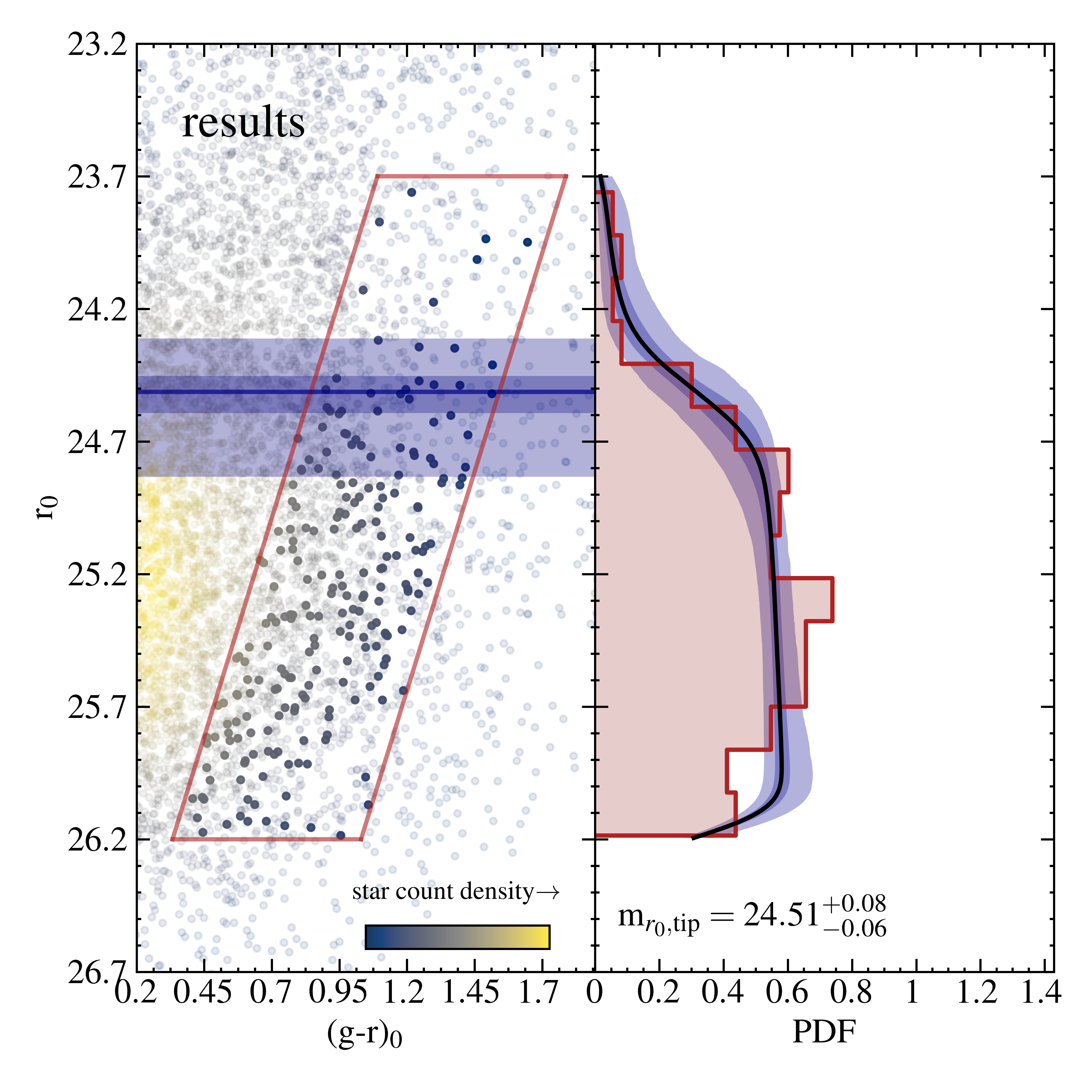}
}
\caption{Measure of the RGB Tip following \citet{bptip24}. Left panel: CMD zoomed on the RGB of KK~153, with the stars colour coded according to the local star count density. The red polygon displays the selection window of the stars used for the detection of the tip, with the additional constraint $R<2.0\arcmin$. The horizontal line marks the position of the tip, while the dark purple and light purple bands mark the $\pm 1\sigma$ and $\pm 3\sigma$ uncertainty regions about the tip. Right panel: luminosity function of the selected KK~153 RGB stars with superimposed the median model (black line) and the $\pm 1\sigma$ and $\pm 3\sigma$ confidence intervals (dark purple and light purple bands, respectively).}
\label{fig:tip}
\end{figure}

To measure the magnitude of the RGB Tip of KK~153 we apply the bayesian inference model described in \citet{bptip24} to the colour-selectd RGB stars lying within $2.0\arcmin$ from the centre of the galaxy. The actual detection is illustrated in Fig.~\ref{fig:tip}, giving $r_0^{TRGB}=24.51^{+0.08}_{-0.06}$.

\citet{bptip24} do not provide a calibration of $M_r^{TRGB}$, since the relatively blue $r$ band is not ideal to trace the tip over a large range of metallicity / RGB colours. However, here we are dealing with low-mass, metal-poor populations, that is a regime where $M_r^{TRGB}$ depends only very weakly on colour \citep{vv124}. Moreover, we derive the distance modulus taking as reference the RGB Tip of the Small Magellanic Cloud (SMC), that has a mean tip colour similar to KK~153
($(g-r)_0^{TRGB}=1.13$, and $(g-r)_0^{TRGB}=1.22$ for KK~153 and the SMC, respectively). Using the same sample and the same technique as \citet{bptip24} we obtained for the SMC $M_r^{TRGB}16.080\pm0.022$. Adopting $(m-M)_0^{SMC}=18.997 \pm0.032$ from \citet{smc20}, we obtain $M_r^{TRGB}=-2.917 \pm 0.039$ for SMC-like RGB populations, in reasonable agreement with the value derived by \citet{vv124} for $(g-r)^{TRGB}_0<1.4$ from theoretical models. 

If we take the SMC as our distance reference, { the distance modulus $(m-M)_0=27.43_{-0.10}^{+0.12}$ is obtained for KK~153, including the uncertainty in the photometric calibration. This corresponds to $D=3.06^{+0.17}_{-0.14}$~Mpc, with uncertainties reaching $^{+0.22}_{-0.20}(^{+0.33}_{-0.31})$~Mpc if an additional uncertainty of 0.1(0.2) magnitudes for the TRGB calibration in r band is hypothesised. 

There is no sound way to estimate the amount of any systematic uncertainty possibly associated to the adopted calibration. As a broad reference we can note that, e.g., a shift of 0.08~mag is required to match the predictions from the PARSEC \citep{bressan2012} models with the adopted measure of $M_{I_E}^{TRGB}$ for the SMC and that the scatter of those models at the colour of the SMC due to an age spread of 9~Gyr is $<0.04$~mag \citep{bptip24}. The exercise performed here by considering 0.1~mag and 0.2~mag additional uncertainty is meant to show that the 1$\sigma$ accuracy of our distance measure remains below $\simeq$10\% also when an unrealistically large uncertainty on the calibration is assumed.} 

\begin{figure}[ht!]
\center{
\includegraphics[width=\columnwidth]{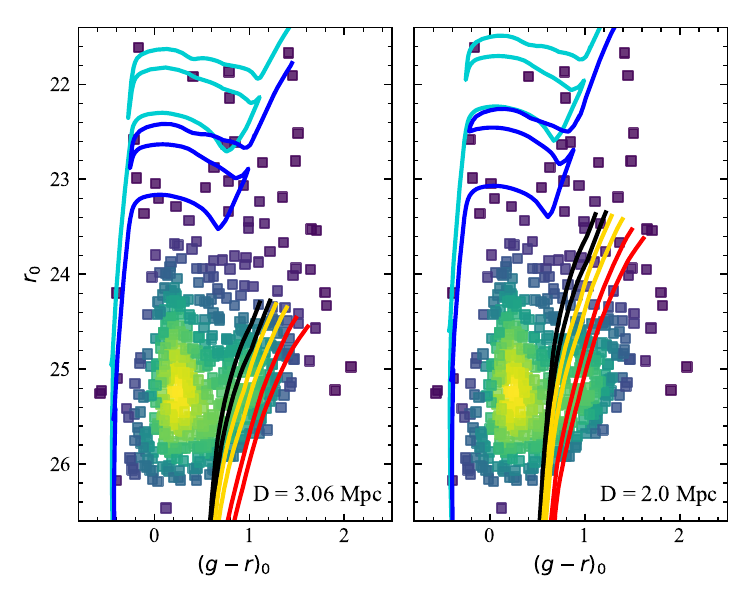}
}
\caption{CMD of KK~153 with isochrones superimposed after correction for the distance 
measured in this paper (${\rm D=3.06}$~Mpc, left panel) and for the distance estimated by
\citet[][${\rm D=2.0}$~Mpc, right panel]{xu25}. The isochrones displaying the RGB phase have metallicity ${\rm [M/H]=-2.0}$ (black), ${\rm [M/H]=-1.5}$ (yellow), and ${\rm[M/H]=-1.0}$ (red). For each metallicity, isochrones of two different ages are plotted, 5.0~Gyr and 10.0~Gyr, where in each pair the older isochrone is redder than the younger one. Two age$\le 100$~Myr isochrones with ${\rm[M/H]=-1.0}$ are also superimposed to both CMDs with the aim of fitting the handful of young KK~153 stars included in our sample. The young isochrones in the left panel have age= 40~Myr (in cyan), and 63~Myr (in blue), in the right panel age= 63~Myr (in cyan), and 100~Myr (in blue).}
\label{fig:iso}
\end{figure}

In Fig.~\ref{fig:iso} we compare the observed CMD of KK~153 with a set of theoretical isochrones from the PARSEC dataset \citep[][obtained with the dedicated web tool\footnote{\url{https://stev.oapd.inaf.it/cgi-bin/cmd}}]{bressan2012} shifted to our newly derived distance (D=3.06~Mpc, left panel) and to the distance inferred by \citet[][D=2.0~Mpc, right panel]{xu25}. The short distance assumption  clearly fails to match the observed RGB. On the other hand, for D=3.06~Mpc the bundle of isochrones spanning the metallicity $-2.0\le [M/H]\le -1.0$ and the age range 5-10~Gyr provides a fully satisfactory fit of the observed RGB. The handful of Main Sequence and Blue Loop stars that we were able to resolve are well matched by $[M/H]=-1.0$ isochrones of age 40~Myr and 63~Myr, suggesting that the star formation was active until very recent epochs in KK~153, possibly still ongoing.

\begin{figure}
\includegraphics[width=0.9\columnwidth]{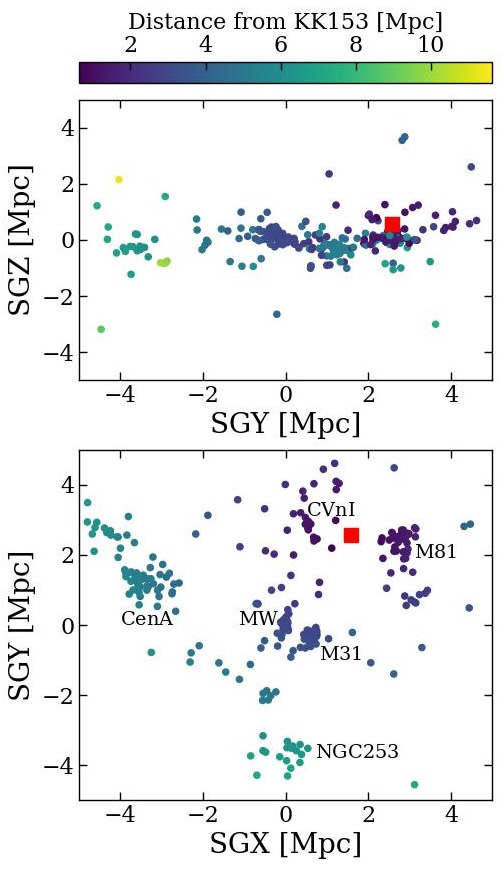}
\caption{Location of KK~153 in cartesian Supergalactic coordinates. Upper Panel: SGY-SGZ projection. Lower panel: SGX-SGY projection. KK~153 is plotted as a large red square while the confirmed dwarfs in the P24 catalogue are plotted as small filled circles color-coded according to their distance from KK~153. The main galaxy groups are labelled in the lower panel.}
\label{fig:supgal}
\end{figure}

The derived distance locates KK~153 significantly farther away than the estimate by \citet{xu25}, clearly beyond the boundaries of the Local Group, and reduces the uncertainties { from more than 50\% to less than 10\%}. Indeed, Fig.~\ref{fig:supgal} shows that  KK~153 is located between the M81 and the CVn~I galaxy groups, and, similarly to the other recently-discovered gas-rich low-stellar-mass dwarfs discussed in \citet{pavo_dist}, is remarkably isolated. The nearest confirmed galaxy in the \citet[][P24 hereafter]{p24} catalogue is UGC~8508 at 630~kpc, itself a pretty isolated dwarf.

\subsection{Stellar mass}
\label{sec:mass}

The distance modulus derived above was used to convert apparent integrated magnitudes into their absolute counterparts $M_g=-10.57 \pm 0.15$ and $M_r=-11.08 \pm 0.15$. $M_V=-10.87\pm 0.20$ was obtained with the photometric transformation
\begin{equation}
    V = g - 0.5784(g-r) -0.0038 
\end{equation}
by Lupton (2005)\footnote{\url{https://www.sdss3.org/dr9/algorithms/sdssUBVRITransform.php\#Lupton2005}}. These absolute magnitudes have been converted in total luminosities in the various passbands assuming the absolute magnitudes of the Sun listed by \citet{willmer18}\footnote{\url{https://mips.as.arizona.edu/~cnaw/sun.html}}.

To convert luminosities into stellar masses we must { refer} to relations providing the stellar mass-to-light (M/L) ratio as a function of the galaxy colour, as done by \citet{xu25}. To somehow account for the systematic uncertainty inherent to these relations and to the models they are derived from we decided to derive $M/L_g$ and $M/L_r$ from various relations from different authors. In particular we used the equations as a function of $(g-r)_0$ colour by \citet{herrmann16}, as done by \citet{xu25}, by \citet{zibetti09}, and the two sets of relations derived from different models by \citet{rc15}, for a total of four different $M/L$ estimates per passband. For the colour of KK~153, $(g-r)_0=0.51$, we obtain $M/L_g$($M/L_r$) values ranging from 1.04(1.01) to 1.43(1.39), implying that the uncertainties tied to the adopted M/L are likely as large as 40\% and possibly larger. We adopt the average of these four values as our reference M/L and their standard deviation as the associated uncertainty, that is $M/L_g=1.18\pm0.17$, and $M/L_r=1.20\pm0.17$.

From these M/L ratios we get $M_{\star}=2.5\pm 0.2\times 10^6
~M_{\sun}$ and $M_{\star}=2.1\pm 0.3\times 10^6
~M_{\sun}$ from $L_g$ and $L_r$, respectively. As our final best estimate of the stellar mass of KK~153 we take the weighted average of these two values, $M_{\star}=2.4\pm 0.2\times 10^6~M_{\sun}$. This is more than five times larger than the estimate by \citet[][$M_{\star}=4.1^{+10.0}_{-2.6}\times 10^5~M_{\sun}$]{xu25}, albeit the statistical difference is less than 2$\sigma$, due to the very large size of the upper error bar  of the Xu et al.'s value. The most important factor contributing to the difference between the two stellar mass estimates is the improved distance (2.35$\times$), then the brighter integrated magnitudes (1.58$\times$), and finally the difference in M/L value, contributing with a 1.46$\times$ factor. The latter difference is due to the bluer colour obtained by \citet{xu25}, $(g-r)_0=0.4$, instead of our $(g-r)_0=0.5$. If our hypothesis of \citet{xu25} measuring the integrated magnitude over an aperture smaller than the one adopted here (Sect.~\ref{sec:apphot}) is correct, it may also explain the difference in colour, since smaller apertures would progressively remove the contribution of red RGB stars to the integrated colour, as these stars populate also the outskirts of the galaxy (and dominate the stellar mass budget), while young blue stars are the prime contributor of light in the central region but are absent outside (and are expected to provide a minor contribution to the mass budget).

\section{Discussion and conclusion}
\label{sec:discu}

We have obtained deep photometry of KK~153, a stellar system recently identified as a local low-stellar-mass gas-rich dwarf galaxy by \citet{xu25}. Based on a very uncertain distance estimate these authors suggested that KK~153 may be a UFD galaxy at the edge of the Local Group, akin to Leo~P \citep{mcquinn15}. However, our newly derived and much more accurate distance, measured from the RGB Tip, relocates the dwarf clearly beyond the Local Group, leading also to a significantly larger stellar mass estimate, significantly beyond the UDF / normal dwarf limit adopted by \citet{xu25}, $M_{\star}\simeq 10^5~M_{\sun}$. Considering the integrated V luminosity $L_V$, KK~153 is $\simeq 4.5$ times brighter than Leo~P, and $\simeq 14$ times brighter than Leo~T (P24). Given the similarity in the stellar content, it can be concluded that KK~153 exceeds these two galaxies by similar factors also in stellar mass.

\begin{figure}[ht!]
\center{
\includegraphics[width=\columnwidth]{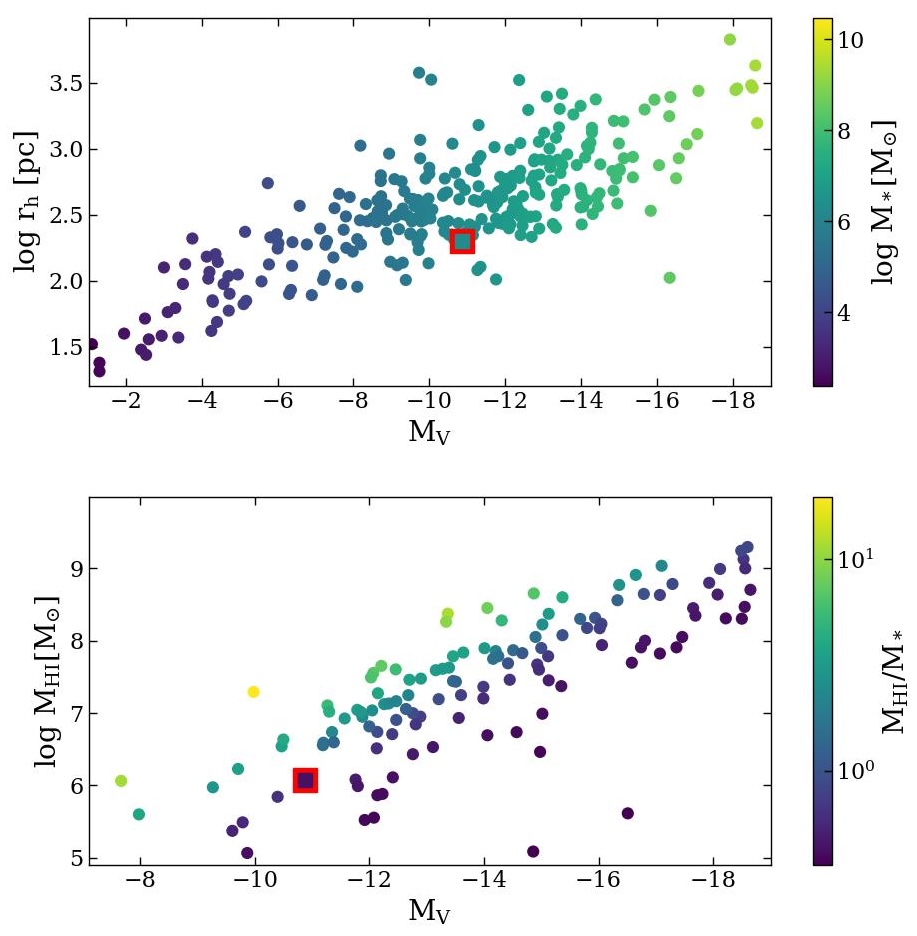}
}
\caption{KK~153 in the context of the confirmed dwarf galaxies in the P24 database. Upper panel: $M_V$ vs. log~$r_h$. Points are colour-coded according to the stellar mass derived from the integrated V luminosity adopting $M/L_V=1.0$ (while in the original P24 list $M/L_V=2.0$ is adopted). Lower panel: $M_V$ vs. log~$M_{\ion{H}{i}}$ for the P24 dwarfs having a valid measure of the \ion{H}{i} mass. The points are colour-coded according to their \ion{H}{i} mass to stellar mass ratio. Also in this case $M/L_V=1.0$ was adopted. In both panels KK~153 { is highlighted with a red square}.}
\label{fig:scale}
\end{figure} 


To properly place KK~153 into the context of local dwarfs, in Fig~\ref{fig:scale} we show its position within the $r_h$ vs. $M_V$ and $M_{\ion{H}{I}}$ vs. $M_V$ relations defined by the confirmed dwarf galaxies listed in the P24 catalogue. KK~153 lies on the high surface brightness edge of the luminosity -- size relation, as typical of isolated dwarfs \citep{pavo_dist}, surrounded by galaxies with similar stellar mass. On the other hand, the lower panel of Fig~\ref{fig:scale} shows that it is indeed among the faintest known dwarfs retaining a measurable amount of \ion{H}{I} in the P24 list, displaying a $M_{\ion{H}{I}}/M_{\star}$ ratio  $\simeq 0.5$, quite typical of dwarf galaxies.
In this plot there are eleven dwarfs fainter than KK~153, four of them are more distant than it, three are part of the NGC~3109 group (Antlia and Antlia~B at D$\simeq 1.3$ Mpc, and Leo~P, at D$\simeq 1.6$ Mpc), and the remaining four are genuine Local Group members, the dwarf irregulars LGS~3 and Acquarius, the transition type dwarf Phoenix, and the gas-rich UFD Leo~T, all having $D\le 1.0$~Mpc. 

In terms of distance, isolation, $M_V$, ellipticity and size KK~153 is remarkably similar to the recently discovered Corvus~A dwarf \citep{corvusA}, that, however, is more gas-rich, having $M_{\ion{H}{I}}/M_{\star}\simeq 2.3$ \citep{pavo_dist}. Both galaxies display a remarkably regular distribution of \ion{H}{I} approximately with the same centre as the star distribution \citep{corvusA,xu25}, in contrast, e.g., with the cases of other similar isolated dwarfs like VV~124(UGC~4879) \citep[][]{vv124} or Pavo \citep{jones25}. This variety of \ion{H}{I} morphologies may suggest that we are observing these galaxies in different phases of their activity, within a cycle that may include star formation - feedback - gas ejection - gas recovery and resettlement to an equilibrium configuration \citep[see, e.g.,][and references therein]{rey22,gutcke22,rey24,mcquinn24,jones25}.  

In a study aimed at exploring the threshold between galaxies quenched by re-ionisation and survivors that were able to form stars after that phase, \citet{gutcke22} performed detailed very high-resolution simulations of five halos in a stellar mass range (at $z=0$) enclosing KK~153, $0.22\times 10^{6}~M_{\sun}\le M_{\star}\le 9.42\times 10^{6}~M_{\sun}$. They found that, while the onset of re-ionisation heavily impacted the star formation of all the considered haloes, three of them were able to re-ignite the star formation after a period of quiescence, albeit at a much lower rate than before. One of these survivors, their `HaloD', displays $z=0$ properties remarkably similar to KK~153 ($M_{\star}=1.6\times 10^6~M_{\sun}$, $r_h = 120$~pc, ongoing star formation). It is intriguing to note that one of the \citet{gutcke22} haloes that was permanently quenched by re-ionisation, 'HaloC', has both a slightly larger halo mass and stellar mass than 'HaloD', highlighting  the delicate interplay of the various factors at play in determining the actual destiny of such small galaxies, not simply total mass \citep[see also][for further discussion and references]{mcquinn24}. The study of systems like KK~153 may help us to get insight into this crucial phase of the evolution of dwarf galaxies, and, indeed they have been a subject of intense research in the last decade
\citep[see, e.g.,][and references therein]{cannon11,cannon15,mcquinn15,sand15_antliab,hir16_leoncino,mcquinn24, pavo_dist,jones25}.

\begin{acknowledgements}

The authors thank an anonymous referee for a constructive and useful report.
MB, FA and DPM acknowledge financial support to this project by INAF, through the PRIN-2023 grant Ob. Fu 1.05.23.05.09 "Dwarf galaxies as probes of the Lambda Cold Dark Matter hierarchical paradigm at the smallest scales" (P.I.: F. Annibali). 
This paper is supported by the Italian Research Center
on High Performance Computing Big Data and Quantum Computing (ICSC),
project funded by European Union - NextGenerationEU - and National Recovery and Resilience Plan (NRRP) - Mission 4 Component 2 within the activities of Spoke 3 (Astrophysics and Cosmos Observations). RP acknowledges the support to this study by the PRIN INAF Mini Grant 2022 (Ob.Fu.
1.05.12.04.02 – CUP C33C22000970005). 
FA acknowledges support by the PRIN INAF Mini Grant 2024 (Ob.Fu.1.05.24.07.02 – CUP C33C24001360005). 

This work is based on LBT data. The LBT is an international collaboration among institutions in the United States, Italy, and Germany. LBT Corporation partners are the University of Arizona on behalf of the Arizona Board of Regents; Istituto Nazionale di Astrofisica, Italy; LBT Beteiligungsgesellschaft, Germany, representing the Max Planck Society, the Leibniz Institute for Astrophysics Potsdam, and Heidelberg University; the Ohio State University, and the Research Corporation, on behalf of the University of Notre Dame, University of Minnesota, and University of Virginia. We acknowledge the support from the LBT-Italian Coordination Facility for the execution of observations, data distribution, and reduction. 

This work made use of SDSS-V data.
Funding for the Sloan Digital Sky Survey V has been provided by the Alfred P. Sloan Foundation, the Heising-Simons Foundation, the National Science Foundation, and the Participating Institutions. SDSS acknowledges support and resources from the Center for High-Performance Computing at the University of Utah. SDSS telescopes are located at Apache Point Observatory, funded by the Astrophysical Research Consortium and operated by New Mexico State University, and at Las Campanas Observatory, operated by the Carnegie Institution for Science. The SDSS web site is \url{www.sdss.org}.

SDSS is managed by the Astrophysical Research Consortium for the Participating Institutions of the SDSS Collaboration, including the Carnegie Institution for Science, Chilean National Time Allocation Committee (CNTAC) ratified researchers, Caltech, the Gotham Participation Group, Harvard University, Heidelberg University, The Flatiron Institute, The Johns Hopkins University, L'Ecole polytechnique f\'{e}d\'{e}rale de Lausanne (EPFL), Leibniz-Institut f\"{u}r Astrophysik Potsdam (AIP), Max-Planck-Institut f\"{u}r Astronomie (MPIA Heidelberg), Max-Planck-Institut f\"{u}r Extraterrestrische Physik (MPE), Nanjing University, National Astronomical Observatories of China (NAOC), New Mexico State University, The Ohio State University, Pennsylvania State University, Smithsonian Astrophysical Observatory, Space Telescope Science Institute (STScI), the Stellar Astrophysics Participation Group, Universidad Nacional Aut\'{o}noma de M\'{e}xico, University of Arizona, University of Colorado Boulder, University of Illinois at Urbana-Champaign, University of Toronto, University of Utah, University of Virginia, Yale University, and Yunnan University.  

This research has made use of the SIMBAD database, operated at CDS, Strasbourg, France. This research has made use of the NASA/IPAC Extragalactic Database (NED), which is operated by the Jet Propulsion Laboratory, California Institute of Technology, under contract with the National Aeronautics and Space Administration. We acknowledge the usage of the HyperLeda database (http://leda.univ-lyon1.fr).

In this analysis we made use of TOPCAT (\url{http://www.starlink.ac.uk/topcat/}, \citealt{Taylor2005}), APT (\url{https://www.aperturephotometry.org} \citealt{apt1,apt2}), DAOPHOT~II ALLFRAME\citep{stet87, allframe}, Sextractor \citep{sextractor}, numpy, astropy, scipy, matplotlib.

\end{acknowledgements}


\bibliographystyle{aa} 
\bibliography{refs} 

@ARTICLE{zibetti09,
       author = {{Zibetti}, Stefano and {Charlot}, St{\'e}phane and {Rix}, Hans-Walter},
        title = "{Resolved stellar mass maps of galaxies - I. Method and implications for global mass estimates}",
      journal = {\mnras},
         year = 2009,
        month = dec,
       volume = {400},
       number = {3},
        pages = {1181-1198},
          doi = {10.1111/j.1365-2966.2009.15528.x},
archivePrefix = {arXiv},
       eprint = {0904.4252},
 primaryClass = {astro-ph.CO},
       adsurl = {https://ui.adsabs.harvard.edu/abs/2009MNRAS.400.1181Z}
}

@ARTICLE{herrmann16,
       author = {{Herrmann}, Kimberly A. and {Hunter}, Deidre A. and {Zhang}, Hong-Xin and {Elmegreen}, Bruce G.},
        title = "{Mass-to-light versus Color Relations for Dwarf Irregular Galaxies}",
      journal = {\aj},
         year = 2016,
        month = dec,
       volume = {152},
       number = {6},
          eid = {177},
        pages = {177},
          doi = {10.3847/0004-6256/152/6/177},
archivePrefix = {arXiv},
       eprint = {1701.00144},
 primaryClass = {astro-ph.GA},
       adsurl = {https://ui.adsabs.harvard.edu/abs/2016AJ....152..177H}
}

@ARTICLE{rc15,
       author = {{Roediger}, Joel C. and {Courteau}, St{\'e}phane},
        title = "{On the uncertainties of stellar mass estimates via colour measurements}",
      journal = {\mnras},
         year = 2015,
        month = sep,
       volume = {452},
       number = {3},
        pages = {3209-3225},
          doi = {10.1093/mnras/stv1499},
archivePrefix = {arXiv},
       eprint = {1507.03016},
 primaryClass = {astro-ph.GA},
       adsurl = {https://ui.adsabs.harvard.edu/abs/2015MNRAS.452.3209R}
}

@ARTICLE{jones23,
       author = {{Jones}, Michael G. and {Mutlu-Pakdil}, Bur{\c{c}}in and {Sand}, David J. and {Donnerstein}, Richard and {Crnojevi{\'c}}, Denija and {Bennet}, Paul and {Fielder}, Catherine E. and {Karunakaran}, Ananthan and {Spekkens}, Kristine and {Strader}, Jay and {Urquhart}, Ryan and {Zaritsky}, Dennis},
        title = "{Pavo: Discovery of a Star-forming Dwarf Galaxy Just Outside the Local Group}",
      journal = {\apjl},
         year = 2023,
        month = nov,
       volume = {957},
       number = {1},
          eid = {L5},
        pages = {L5},
          doi = {10.3847/2041-8213/ad0130},
archivePrefix = {arXiv},
       eprint = {2310.01478},
 primaryClass = {astro-ph.GA},
       adsurl = {https://ui.adsabs.harvard.edu/abs/2023ApJ...957L...5J}
}

@ARTICLE{giova15,
       author = {{Giovanelli}, Riccardo and {Haynes}, Martha P.},
        title = "{Extragalactic HI surveys}",
      journal = {\aapr},
         year = 2015,
        month = dec,
       volume = {24},
          eid = {1},
        pages = {1},
          doi = {10.1007/s00159-015-0085-3},
archivePrefix = {arXiv},
       eprint = {1510.04660},
 primaryClass = {astro-ph.GA},
       adsurl = {https://ui.adsabs.harvard.edu/abs/2015A&ARv..24....1G}
}

@ARTICLE{leisman21,
       author = {{Leisman}, Lukas and {Rhode}, Katherine L. and {Ball}, Catherine and {Pagel}, Hannah J. and {Cannon}, John M. and {Salzer}, John J. and {Janowiecki}, Steven and {Janesh}, William F. and {J{\'o}zsa}, Gyula I.~G. and {Giovanelli}, Riccardo and {Haynes}, Martha P. and {Adams}, Elizabeth A.~K. and {Gray}, Laurin and {Smith}, Nicholas J.},
        title = "{The ALFALFA Almost Dark Galaxy AGC 229101: A 2 Billion Solar Mass H I Cloud with a Very Low Surface Brightness Optical Counterpart}",
      journal = {\aj},
         year = 2021,
        month = dec,
       volume = {162},
       number = {6},
          eid = {274},
        pages = {274},
          doi = {10.3847/1538-3881/ac2a38},
archivePrefix = {arXiv},
       eprint = {2109.12139},
 primaryClass = {astro-ph.GA},
       adsurl = {https://ui.adsabs.harvard.edu/abs/2021AJ....162..274L}
}

@ARTICLE{leisman17,
       author = {{Leisman}, Lukas and {Haynes}, Martha P. and {Janowiecki}, Steven and {Hallenbeck}, Gregory and {J{\'o}zsa}, Gyula and {Giovanelli}, Riccardo and {Adams}, Elizabeth A.~K. and {Bernal Neira}, David and {Cannon}, John M. and {Janesh}, William F. and {Rhode}, Katherine L. and {Salzer}, John J.},
        title = "{(Almost) Dark Galaxies in the ALFALFA Survey: Isolated H I-bearing Ultra-diffuse Galaxies}",
      journal = {\apj},
         year = 2017,
        month = jun,
       volume = {842},
       number = {2},
          eid = {133},
        pages = {133},
          doi = {10.3847/1538-4357/aa7575},
archivePrefix = {arXiv},
       eprint = {1703.05293},
 primaryClass = {astro-ph.GA},
       adsurl = {https://ui.adsabs.harvard.edu/abs/2017ApJ...842..133L}
}

@ARTICLE{anand25,
       author = {{Anand}, Gagandeep S. and {Ben{\'\i}tez-Llambay}, Alejandro and {Beaton}, Rachael and {Fox}, Andrew J. and {Navarro}, Julio F. and {D'Onghia}, Elena},
        title = "{The First RELHIC? Cloud-9 is a Starless Gas Cloud}",
      journal = {arXiv e-prints},
         year = 2025,
        month = aug,
          eid = {arXiv:2508.20157},
        pages = {arXiv:2508.20157},
          doi = {10.48550/arXiv.2508.20157},
archivePrefix = {arXiv},
       eprint = {2508.20157},
 primaryClass = {astro-ph.GA},
       adsurl = {https://ui.adsabs.harvard.edu/abs/2025arXiv250820157A}
}

@ARTICLE{rey24,
       author = {{Rey}, Martin P. and {Orkney}, Matthew D.~A. and {Read}, Justin I. and {Das}, Payel and {Agertz}, Oscar and {Pontzen}, Andrew and {Ponomareva}, Anastasia A. and {Kim}, Stacy Y. and {McClymont}, William},
        title = "{EDGE - Dark matter or astrophysics? Breaking dark matter heating degeneracies with H I rotation in faint dwarf galaxies}",
      journal = {\mnras},
         year = 2024,
        month = apr,
       volume = {529},
       number = {3},
        pages = {2379-2398},
          doi = {10.1093/mnras/stae718},
archivePrefix = {arXiv},
       eprint = {2309.00041},
 primaryClass = {astro-ph.GA},
       adsurl = {https://ui.adsabs.harvard.edu/abs/2024MNRAS.529.2379R}
}

@ARTICLE{rey22,
       author = {{Rey}, Martin P. and {Pontzen}, Andrew and {Agertz}, Oscar and {Orkney}, Matthew D.~A. and {Read}, Justin I. and {Saintonge}, Am{\'e}lie and {Kim}, Stacy Y. and {Das}, Payel},
        title = "{EDGE: What shapes the relationship between H I and stellar observables in faint dwarf galaxies?}",
      journal = {\mnras},
         year = 2022,
        month = apr,
       volume = {511},
       number = {4},
        pages = {5672-5681},
          doi = {10.1093/mnras/stac502},
archivePrefix = {arXiv},
       eprint = {2112.03280},
 primaryClass = {astro-ph.GA},
       adsurl = {https://ui.adsabs.harvard.edu/abs/2022MNRAS.511.5672R}
}

@ARTICLE{corvusA,
       author = {{Jones}, Michael G. and {Sand}, David J. and {Mutlu-Pakdil}, Bur{\c{c}}in and {Fielder}, Catherine E. and {Crnojevi{\'c}}, Denija and {Bennet}, Paul and {Spekkens}, Kristine and {Donnerstein}, Richard and {Doliva-Dolinsky}, Amandine and {Karunakaran}, Ananthan and {Strader}, Jay and {Zaritsky}, Dennis},
        title = "{Corvus A: A Low-mass, Isolated Galaxy at 3.5 Mpc}",
      journal = {\apjl},
         year = 2024,
        month = aug,
       volume = {971},
       number = {2},
          eid = {L37},
        pages = {L37},
          doi = {10.3847/2041-8213/ad676e},
archivePrefix = {arXiv},
       eprint = {2407.03393},
 primaryClass = {astro-ph.GA},
       adsurl = {https://ui.adsabs.harvard.edu/abs/2024ApJ...971L..37J}
}

@ARTICLE{mcquinn24,
       author = {{McQuinn}, Kristen B.~W. and {Newman}, Max J.~B. and {Skillman}, Evan D. and {Telford}, O. Grace and {Brooks}, Alyson and {Adams}, Elizabeth A.~K. and {Berg}, Danielle A. and {Boyer}, Martha L. and {Cannon}, John M. and {Dolphin}, Andrew E. and {Pahl}, Anthony J. and {Rhode}, Katherine L. and {Salzer}, John J. and {Cohen}, Roger E. and {Goldman}, Steve R.},
        title = "{The Ancient Star Formation History of the Extremely Low-mass Galaxy Leo P: An Emerging Trend of a Post-reionization Pause in Star Formation}",
      journal = {\apj},
         year = 2024,
        month = nov,
       volume = {976},
       number = {1},
          eid = {60},
        pages = {60},
          doi = {10.3847/1538-4357/ad8158},
archivePrefix = {arXiv},
       eprint = {2409.19050},
 primaryClass = {astro-ph.GA},
       adsurl = {https://ui.adsabs.harvard.edu/abs/2024ApJ...976...60M}
}

@ARTICLE{cannon11,
       author = {{Cannon}, John M. and {Giovanelli}, Riccardo and {Haynes}, Martha P. and {Janowiecki}, Steven and {Parker}, Angela and {Salzer}, John J. and {Adams}, Elizabeth A.~K. and {Engstrom}, Eric and {Huang}, Shan and {McQuinn}, Kristen B.~W. and {Ott}, J{\"u}rgen and {Saintonge}, Am{\'e}lie and {Skillman}, Evan D. and {Allan}, John and {Erny}, Grace and {Fliss}, Palmer and {Smith}, AnnaLeigh},
        title = "{The Survey of H I in Extremely Low-mass Dwarfs (SHIELD)}",
      journal = {\apjl},
         year = 2011,
        month = sep,
       volume = {739},
       number = {1},
          eid = {L22},
        pages = {L22},
          doi = {10.1088/2041-8205/739/1/L22},
archivePrefix = {arXiv},
       eprint = {1105.4505},
 primaryClass = {astro-ph.CO},
       adsurl = {https://ui.adsabs.harvard.edu/abs/2011ApJ...739L..22C}
}

@ARTICLE{hir16_leoncino,
       author = {{Hirschauer}, Alec S. and {Salzer}, John J. and {Skillman}, Evan D. and {Berg}, Danielle and {McQuinn}, Kristen B.~W. and {Cannon}, John M. and {Gordon}, Alex J.~R. and {Haynes}, Martha P. and {Giovanelli}, Riccardo and {Adams}, Elizabeth A.~K. and {Janowiecki}, Steven and {Rhode}, Katherine L. and {Pogge}, Richard W. and {Croxall}, Kevin V. and {Aver}, Erik},
        title = "{ALFALFA Discovery of the Most Metal-poor Gas-rich Galaxy Known: AGC 198691}",
      journal = {\apj},
         year = 2016,
        month = may,
       volume = {822},
       number = {2},
          eid = {108},
        pages = {108},
          doi = {10.3847/0004-637X/822/2/108},
archivePrefix = {arXiv},
       eprint = {1603.03798},
 primaryClass = {astro-ph.GA},
       adsurl = {https://ui.adsabs.harvard.edu/abs/2016ApJ...822..108H}
}

@ARTICLE{sand15_antliab,
       author = {{Sand}, D.~J. and {Spekkens}, K. and {Crnojevi{\'c}}, D. and {Hargis}, J.~R. and {Willman}, B. and {Strader}, J. and {Grillmair}, C.~J.},
        title = "{Antlia B: A Faint Dwarf Galaxy Member of the NGC 3109 Association}",
      journal = {\apjl},
         year = 2015,
        month = oct,
       volume = {812},
       number = {1},
          eid = {L13},
        pages = {L13},
          doi = {10.1088/2041-8205/812/1/L13},
archivePrefix = {arXiv},
       eprint = {1508.01800},
 primaryClass = {astro-ph.GA},
       adsurl = {https://ui.adsabs.harvard.edu/abs/2015ApJ...812L..13S}
}

@ARTICLE{willmer18,
       author = {{Willmer}, Christopher N.~A.},
        title = "{The Absolute Magnitude of the Sun in Several Filters}",
      journal = {\apjs},
         year = 2018,
        month = jun,
       volume = {236},
       number = {2},
          eid = {47},
        pages = {47},
          doi = {10.3847/1538-4365/aabfdf},
archivePrefix = {arXiv},
       eprint = {1804.07788},
 primaryClass = {astro-ph.SR},
       adsurl = {https://ui.adsabs.harvard.edu/abs/2018ApJS..236...47W}
}

@ARTICLE{smc20,
       author = {{Graczyk}, Dariusz and {Pietrzy{\'n}ski}, Grzegorz and {Thompson}, Ian B. and {Gieren}, Wolfgang and {Zgirski}, Bart{\l}omiej and {Villanova}, Sandro and {G{\'o}rski}, Marek and {Wielg{\'o}rski}, Piotr and {Karczmarek}, Paulina and {Narloch}, Weronika and {Pilecki}, Bogumi{\l} and {Taormina}, Monica and {Smolec}, Rados{\l}aw and {Suchomska}, Ksenia and {Gallenne}, Alexandre and {Nardetto}, Nicolas and {Storm}, Jesper and {Kudritzki}, Rolf-Peter and {Ka{\l}uszy{\'n}ski}, Miko{\l}aj and {Pych}, Wojciech},
        title = "{A Distance Determination to the Small Magellanic Cloud with an Accuracy of Better than Two Percent Based on Late-type Eclipsing Binary Stars}",
      journal = {\apj},
         year = 2020,
        month = nov,
       volume = {904},
       number = {1},
          eid = {13},
        pages = {13},
          doi = {10.3847/1538-4357/abbb2b},
archivePrefix = {arXiv},
       eprint = {2010.08754},
 primaryClass = {astro-ph.GA},
       adsurl = {https://ui.adsabs.harvard.edu/abs/2020ApJ...904...13G}
}

@ARTICLE{perina09,
       author = {{Perina}, S. and {Barmby}, P. and {Beasley}, M.~A. and {Bellazzini}, M. and {Brodie}, J.~P. and {Burstein}, D. and {Cohen}, J.~G. and {Federici}, L. and {Fusi Pecci}, F. and {Galleti}, S. and {Hodge}, P.~W. and {Huchra}, J.~P. and {Kissler-Patig}, M. and {Puzia}, T.~H. and {Strader}, J.},
        title = "{An HST/WFPC2 survey of bright young clusters in M31. I. VdB0, a massive star cluster seen at t ≃ 25 Myr}",
      journal = {\aap},
         year = 2009,
        month = feb,
       volume = {494},
       number = {3},
        pages = {933-948},
          doi = {10.1051/0004-6361:200810725},
archivePrefix = {arXiv},
       eprint = {0812.1668},
 primaryClass = {astro-ph},
       adsurl = {https://ui.adsabs.harvard.edu/abs/2009A&A...494..933P}
}

@ARTICLE{hyperleda,
   author = {{Makarov}, D. and {Prugniel}, P. and {Terekhova}, N. and {Courtois}, H. and 
	{Vauglin}, I.},
    title = "{HyperLEDA. III. The catalogue of extragalactic distances}",
  journal = {\aap},
     year = 2014,
    month = oct,
   volume = 570,
      eid = {A13},
    pages = {A13},
      doi = {10.1051/0004-6361/201423496},
   adsurl = {http://adsabs.harvard.edu/abs/2014A%26A...570A..13M}
}

@ARTICLE{gutcke22,
       author = {{Gutcke}, Thales A. and {Pfrommer}, Christoph and {Bryan}, Greg L. and {Pakmor}, R{\"u}diger and {Springel}, Volker and {Naab}, Thorsten},
        title = "{LYRA. III. The Smallest Reionization Survivors}",
      journal = {\apj},
         year = 2022,
        month = dec,
       volume = {941},
       number = {2},
          eid = {120},
        pages = {120},
          doi = {10.3847/1538-4357/aca1b4},
archivePrefix = {arXiv},
       eprint = {2209.03366},
 primaryClass = {astro-ph.GA},
       adsurl = {https://ui.adsabs.harvard.edu/abs/2022ApJ...941..120G}
}

@ARTICLE{irwin07,
       author = {{Irwin}, M.~J. and {Belokurov}, V. and {Evans}, N.~W. and {Ryan-Weber}, E.~V. and {de Jong}, J.~T.~A. and {Koposov}, S. and {Zucker}, D.~B. and {Hodgkin}, S.~T. and {Gilmore}, G. and {Prema}, P. and {Hebb}, L. and {Begum}, A. and {Fellhauer}, M. and {Hewett}, P.~C. and {Kennicutt}, Jr., R.~C. and {Wilkinson}, M.~I. and {Bramich}, D.~M. and {Vidrih}, S. and {Rix}, H. -W. and {Beers}, T.~C. and {Barentine}, J.~C. and {Brewington}, H. and {Harvanek}, M. and {Krzesinski}, J. and {Long}, D. and {Nitta}, A. and {Snedden}, S.~A.},
        title = "{Discovery of an Unusual Dwarf Galaxy in the Outskirts of the Milky Way}",
      journal = {\apjl},
         year = 2007,
        month = feb,
       volume = {656},
       number = {1},
        pages = {L13-L16},
          doi = {10.1086/512183},
archivePrefix = {arXiv},
       eprint = {astro-ph/0701154},
 primaryClass = {astro-ph},
       adsurl = {https://ui.adsabs.harvard.edu/abs/2007ApJ...656L..13I}
}

@ARTICLE{pavo_dist,
       author = {{Mutlu-Pakdil}, Bur{\c{c}}in and {Jones}, Michael G. and {Sand}, David J. and {Crnojevi{\'c}}, Denija and {Herron}, Kai and {Strader}, Jay and {Zaritsky}, Dennis and {Bennet}, Paul and {Drlica-Wagner}, Alex and {Casey}, Quinn O. and {Doliva-Dolinsky}, Amandine and {Donnerstein}, Richard and {Fielder}, Catherine E. and {Hunter}, Laura C. and {Peter}, Annika H.~G. and {Prabhu}, Deepthi S. and {Spekkens}, Kristine},
        title = "{Hubble Space Telescope Imaging of Three Isolated Faint Dwarf Galaxies Beyond the Local Group: Pavo, Corvus A, and Kamino}",
      journal = {arXiv e-prints},
         year = 2025,
        month = sep,
          eid = {arXiv:2509.16307},
        pages = {arXiv:2509.16307},
          doi = {10.48550/arXiv.2509.16307},
archivePrefix = {arXiv},
       eprint = {2509.16307},
 primaryClass = {astro-ph.GA},
       adsurl = {https://ui.adsabs.harvard.edu/abs/2025arXiv250916307M}
}

@ARTICLE{jones25,
       author = {{Jones}, Michael G. and {Rey}, Martin P. and {Sand}, David J. and {Spekkens}, Kristine and {Mutlu-Pakdil}, Bur{\c{c}}in and {Adams}, Elizabeth A.~K. and {Bennet}, Paul and {Crnojevi{\'c}}, Denija and {Doliva-Dolinsky}, Amandine and {Donnerstein}, Richard and {Fielder}, Catherine E. and {Healy}, Julia and {Hunter}, Laura C. and {Karunakaran}, Ananthan and {Prabhu}, Deepthi S. and {Zaritsky}, Dennis},
        title = "{Pavo: Stellar Feedback in Action in a Low-mass Dwarf Galaxy}",
      journal = {\apj},
         year = 2025,
        month = sep,
       volume = {990},
       number = {2},
          eid = {164},
        pages = {164},
          doi = {10.3847/1538-4357/adf6ab},
archivePrefix = {arXiv},
       eprint = {2506.06424},
 primaryClass = {astro-ph.GA},
       adsurl = {https://ui.adsabs.harvard.edu/abs/2025ApJ...990..164J}
}

@ARTICLE{zhang24,
        author = {{Zhang}, Chuan-Peng and {Zhu}, Ming and {Jiang}, Peng and {Cheng}, Cheng and {Wang}, Jing and {Wang}, Jie and {Xu}, Jin-Long and {Liu}, Xiao-Lan and {Yu}, Nai-Ping and {Qian}, Lei and {Yu}, Haiyang and {Ai}, Mei and {Jing}, Yingjie and {Xu}, Chen and {Liu}, Ziming and {Guan}, Xin and {Sun}, Chun and {Yang}, Qingliang and {Huang}, Menglin and {Hao}, Qiaoli and {FAST Collaboration}},
        title = "{The FAST all sky H I survey (FASHI): The first release of catalog}",
      journal = {Science China Physics, Mechanics, and Astronomy},
         year = 2024,
        month = jan,
       volume = {67},
       number = {1},
          eid = {219511},
        pages = {219511},
          doi = {10.1007/s11433-023-2219-7},
archivePrefix = {arXiv},
       eprint = {2312.06097},
 primaryClass = {astro-ph.GA},
       adsurl = {https://ui.adsabs.harvard.edu/abs/2024SCPMA..6719511Z}
}

@ARTICLE{mcquinn15,
       author = {{McQuinn}, Kristen B.~W. and {Skillman}, Evan D. and {Dolphin}, Andrew and {Cannon}, John M. and {Salzer}, John J. and {Rhode}, Katherine L. and {Adams}, Elizabeth A.~K. and {Berg}, Danielle and {Giovanelli}, Riccardo and {Girardi}, L{\'e}o and {Haynes}, Martha P.},
        title = "{Leo P: An Unquenched Very Low-mass Galaxy}",
      journal = {\apj},
         year = 2015,
        month = oct,
       volume = {812},
       number = {2},
          eid = {158},
        pages = {158},
          doi = {10.1088/0004-637X/812/2/158},
archivePrefix = {arXiv},
       eprint = {1506.05495},
 primaryClass = {astro-ph.GA},
       adsurl = {https://ui.adsabs.harvard.edu/abs/2015ApJ...812..158M}
}

@ARTICLE{giova13,
       author = {{Giovanelli}, Riccardo and {Haynes}, Martha P. and {Adams}, Elizabeth A.~K. and {Cannon}, John M. and {Rhode}, Katherine L. and {Salzer}, John J. and {Skillman}, Evan D. and {Bernstein-Cooper}, Elijah Z. and {McQuinn}, Kristen B.~W.},
        title = "{ALFALFA Discovery of the Nearby Gas-rich Dwarf Galaxy Leo P. I. H I Observations}",
      journal = {\aj},
         year = 2013,
        month = jul,
       volume = {146},
       number = {1},
          eid = {15},
        pages = {15},
          doi = {10.1088/0004-6256/146/1/15},
archivePrefix = {arXiv},
       eprint = {1305.0272},
 primaryClass = {astro-ph.CO},
       adsurl = {https://ui.adsabs.harvard.edu/abs/2013AJ....146...15G}
}

@ARTICLE{skill13,
       author = {{Skillman}, Evan D. and {Salzer}, John J. and {Berg}, Danielle A. and {Pogge}, Richard W. and {Haurberg}, Nathalie C. and {Cannon}, John M. and {Aver}, Erik and {Olive}, Keith A. and {Giovanelli}, Riccardo and {Haynes}, Martha P. and {Adams}, Elizabeth A.~K. and {McQuinn}, Kristen B.~W. and {Rhode}, Katherine L.},
        title = "{ALFALFA Discovery of the nearby Gas-rich Dwarf Galaxy Leo P. III. An Extremely Metal Deficient Galaxy}",
      journal = {\aj},
         year = 2013,
        month = jul,
       volume = {146},
       number = {1},
          eid = {3},
        pages = {3},
          doi = {10.1088/0004-6256/146/1/3},
archivePrefix = {arXiv},
       eprint = {1305.0277},
 primaryClass = {astro-ph.CO},
       adsurl = {https://ui.adsabs.harvard.edu/abs/2013AJ....146....3S}
}

@ARTICLE{mic15,
       author = {{Bellazzini}, M. and {Magrini}, L. and {Mucciarelli}, A. and {Beccari}, G. and {Ibata}, R. and {Battaglia}, G. and {Martin}, N. and {Testa}, V. and {Fumana}, M. and {Marchetti}, A. and {Correnti}, M. and {Fraternali}, F.},
        title = "{H II Regions Within a Compact High Velocity Cloud. A Nearly Starless Dwarf Galaxy?}",
      journal = {\apjl},
         year = 2015,
        month = feb,
       volume = {800},
       number = {1},
          eid = {L15},
        pages = {L15},
          doi = {10.1088/2041-8205/800/1/L15},
       adsurl = {https://ui.adsabs.harvard.edu/abs/2015ApJ...800L..15B}
}

@ARTICLE{sand15,
       author = {{Sand}, D.~J. and {Crnojevi{\'c}}, D. and {Bennet}, P. and {Willman}, B. and {Hargis}, J. and {Strader}, J. and {Olszewski}, E. and {Tollerud}, E.~J. and {Simon}, J.~D. and {Caldwell}, N. and {Guhathakurta}, P. and {James}, B.~L. and {Koposov}, S. and {McLeod}, B. and {Morrell}, N. and {Peacock}, M. and {Salinas}, R. and {Seth}, A.~C. and {Stark}, D.~P. and {Toloba}, E.},
        title = "{A Comprehensive Archival Search for Counterparts to Ultra-compact High-Velocity Clouds: Five Local Volume Dwarf Galaxies}",
      journal = {\apj},
         year = 2015,
        month = jun,
       volume = {806},
       number = {1},
          eid = {95},
        pages = {95},
          doi = {10.1088/0004-637X/806/1/95},
archivePrefix = {arXiv},
       eprint = {1503.00720},
 primaryClass = {astro-ph.GA},
       adsurl = {https://ui.adsabs.harvard.edu/abs/2015ApJ...806...95S}
}

@ARTICLE{cannon15,
       author = {{Cannon}, John M. and {Martinkus}, Charlotte P. and {Leisman}, Lukas and {Haynes}, Martha P. and {Adams}, Elizabeth A.~K. and {Giovanelli}, Riccardo and {Hallenbeck}, Gregory and {Janowiecki}, Steven and {Jones}, Michael and {J{\'o}zsa}, Gyula I.~G. and {Koopmann}, Rebecca A. and {Nichols}, Nathan and {Papastergis}, Emmanouil and {Rhode}, Katherine L. and {Salzer}, John J. and {Troischt}, Parker},
        title = "{The Alfalfa {\textquotedblleft}Almost Darks{\textquotedblright} Campaign: Pilot VLA HI Observations of Five High Mass-To-Light Ratio Systems}",
      journal = {\aj},
         year = 2015,
        month = feb,
       volume = {149},
       number = {2},
          eid = {72},
        pages = {72},
          doi = {10.1088/0004-6256/149/2/72},
archivePrefix = {arXiv},
       eprint = {1412.3018},
 primaryClass = {astro-ph.GA},
       adsurl = {https://ui.adsabs.harvard.edu/abs/2015AJ....149...72C}
}

@ARTICLE{saul14,
       author = {{Saul}, Destry R. and {Peek}, J.~E.~G. and {Putman}, M.~E.},
        title = "{Dust-to-gas ratios of the GALFA-H I Compact Cloud Catalog}",
      journal = {\mnras},
         year = 2014,
        month = jul,
       volume = {441},
       number = {3},
        pages = {2266-2272},
          doi = {10.1093/mnras/stu498},
archivePrefix = {arXiv},
       eprint = {1403.4617},
 primaryClass = {astro-ph.GA},
       adsurl = {https://ui.adsabs.harvard.edu/abs/2014MNRAS.441.2266S}
}

@ARTICLE{adams13,
       author = {{Adams}, Elizabeth A.~K. and {Giovanelli}, Riccardo and {Haynes}, Martha P.},
        title = "{A Catalog of Ultra-compact High Velocity Clouds from the ALFALFA Survey: Local Group Galaxy Candidates?}",
      journal = {\apj},
         year = 2013,
        month = may,
       volume = {768},
       number = {1},
          eid = {77},
        pages = {77},
          doi = {10.1088/0004-637X/768/1/77},
archivePrefix = {arXiv},
       eprint = {1303.6967},
 primaryClass = {astro-ph.CO},
       adsurl = {https://ui.adsabs.harvard.edu/abs/2013ApJ...768...77A}
}

@ARTICLE{belo13,
       author = {{Belokurov}, Vasily},
        title = "{Galactic Archaeology: The dwarfs that survived and perished}",
      journal = {\nar},
         year = 2013,
        month = sep,
       volume = {57},
       number = {3-4},
        pages = {100-121},
          doi = {10.1016/j.newar.2013.07.001},
archivePrefix = {arXiv},
       eprint = {1307.0041},
 primaryClass = {astro-ph.GA},
       adsurl = {https://ui.adsabs.harvard.edu/abs/2013NewAR..57..100B}
}

@ARTICLE{simon19,
       author = {{Simon}, Joshua D.},
        title = "{The Faintest Dwarf Galaxies}",
      journal = {\araa},
         year = 2019,
        month = aug,
       volume = {57},
        pages = {375-415},
          doi = {10.1146/annurev-astro-091918-104453},
archivePrefix = {arXiv},
       eprint = {1901.05465},
 primaryClass = {astro-ph.GA},
       adsurl = {https://ui.adsabs.harvard.edu/abs/2019ARA&A..57..375S}
}

@ARTICLE{xu25,
       author = {{Xu}, Jin-Long and {Zhu}, Ming and {Yu}, Nai-Ping and {Zhang}, Chuan-Peng and {Liu}, Xiao-Lan and {Ai}, Mei and {Jiang}, Peng},
        title = "{FAST Discovery of a Gas-rich and Ultrafaint Dwarf Galaxy: KK153}",
      journal = {\apjl},
         year = 2025,
        month = apr,
       volume = {982},
       number = {2},
          eid = {L36},
        pages = {L36},
          doi = {10.3847/2041-8213/adbe7e},
archivePrefix = {arXiv},
       eprint = {2503.08999},
 primaryClass = {astro-ph.GA},
       adsurl = {https://ui.adsabs.harvard.edu/abs/2025ApJ...982L..36X}
}

@ARTICLE{bptip24,
       author = {{Bellazzini}, M. and {Pascale}, R.},
        title = "{The red giant branch tip in the SDSS, PS1, JWST, NGRST, and Euclid photometric systems: Calibration in optical passbands using Gaia DR3 synthetic photometry}",
      journal = {\aap},
         year = 2024,
        month = nov,
       volume = {691},
          eid = {A42},
        pages = {A42},
          doi = {10.1051/0004-6361/202449575},
archivePrefix = {arXiv},
       eprint = {2406.04781},
 primaryClass = {astro-ph.GA},
       adsurl = {https://ui.adsabs.harvard.edu/abs/2024A&A...691A..42B}
}

@ARTICLE{secco,
       author = {{Bellazzini}, M. and {Beccari}, G. and {Battaglia}, G. and {Martin}, N. and {Testa}, V. and {Ibata}, R. and {Correnti}, M. and {Cusano}, F. and {Sani}, E.},
        title = "{The StEllar Counterparts of COmpact high velocity clouds (SECCO) survey. I. Photos of ghosts}",
      journal = {\aap},
         year = 2015,
        month = mar,
       volume = {575},
          eid = {A126},
        pages = {A126},
          doi = {10.1051/0004-6361/201425262},
archivePrefix = {arXiv},
       eprint = {1412.5857},
 primaryClass = {astro-ph.GA},
       adsurl = {https://ui.adsabs.harvard.edu/abs/2015A&A...575A.126B}
}

@ARTICLE{lee93,
       author = {{Lee}, Myung Gyoon and {Freedman}, Wendy L. and {Madore}, Barry F.},
        title = "{The Tip of the Red Giant Branch as a Distance Indicator for Resolved Galaxies}",
      journal = {\apj},
         year = 1993,
        month = nov,
       volume = {417},
        pages = {553},
          doi = {10.1086/173334},
       adsurl = {https://ui.adsabs.harvard.edu/abs/1993ApJ...417..553L}
}

@ARTICLE{salacas98,
       author = {{Salaris}, Maurizio and {Cassisi}, Santi},
        title = "{A new analysis of the red giant branch `tip' distance scale and the value of the Hubble constant}",
      journal = {\mnras},
         year = 1998,
        month = jul,
       volume = {298},
       number = {1},
        pages = {166-178},
          doi = {10.1046/j.1365-8711.1998.01598.x},
archivePrefix = {arXiv},
       eprint = {astro-ph/9803103},
 primaryClass = {astro-ph},
       adsurl = {https://ui.adsabs.harvard.edu/abs/1998MNRAS.298..166S}
}

@ARTICLE{madore20,
       author = {{Madore}, Barry F. and {Freedman}, Wendy L.},
        title = "{Mathematical Underpinnings of the Multiwavelength Structure of the Tip of the Red Giant Branch}",
      journal = {\aj},
         year = 2020,
        month = oct,
       volume = {160},
       number = {4},
          eid = {170},
        pages = {170},
          doi = {10.3847/1538-3881/abab9a},
archivePrefix = {arXiv},
       eprint = {2008.00341},
 primaryClass = {astro-ph.GA},
       adsurl = {https://ui.adsabs.harvard.edu/abs/2020AJ....160..170M}
}

@ARTICLE{sacchi24,
       author = {{Sacchi}, Elena and {Bellazzini}, Michele and {Annibali}, Francesca and {Tosi}, Monica and {Beccari}, Giacomo and {Cannon}, John M. and {Hunter}, Laura C. and {Paris}, Diego and {Roychowdhury}, Sambit and {Schisgal}, Lila and {van Zee}, Liese and {Cignoni}, Michele and {Cusano}, Felice and {de Jong}, Roelof S. and {Hunt}, Leslie and {Pascale}, Raffaele},
        title = "{The Smallest Scale of Hierarchy Survey (SSH): III. Dwarf-dwarf satellite merging phenomena in the low-mass regime}",
      journal = {\aap},
         year = 2024,
        month = nov,
       volume = {691},
          eid = {A65},
        pages = {A65},
          doi = {10.1051/0004-6361/202450106},
archivePrefix = {arXiv},
       eprint = {2406.01683},
 primaryClass = {astro-ph.GA},
       adsurl = {https://ui.adsabs.harvard.edu/abs/2024A&A...691A..65S}
}

@ARTICLE{ssh_pap1,
       author = {{Annibali}, F. and {Beccari}, G. and {Bellazzini}, M. and {Tosi}, M. and {Cusano}, F. and {Paris}, D. and {Cignoni}, M. and {Ciotti}, L. and {Nipoti}, C. and {Sacchi}, E.},
        title = "{The Smallest Scale of Hierarchy survey (SSH) - I. Survey description}",
      journal = {\mnras},
         year = 2020,
        month = feb,
       volume = {491},
       number = {4},
        pages = {5101-5125},
          doi = {10.1093/mnras/stz3185},
archivePrefix = {arXiv},
       eprint = {1911.08543},
 primaryClass = {astro-ph.GA},
       adsurl = {https://ui.adsabs.harvard.edu/abs/2020MNRAS.491.5101A}
}

@ARTICLE{vv124,
       author = {{Bellazzini}, M. and {Perina}, S. and {Galleti}, S. and {Oosterloo}, T.},
        title = "{HST-ACS photometry of the isolated dwarf galaxy VV124=UGC 4879. Detection of the blue horizontal branch and identification of two young star clusters}",
      journal = {\aap},
         year = 2011,
        month = sep,
       volume = {533},
          eid = {A37},
        pages = {A37},
          doi = {10.1051/0004-6361/201117275},
archivePrefix = {arXiv},
       eprint = {1107.2556},
 primaryClass = {astro-ph.CO},
       adsurl = {https://ui.adsabs.harvard.edu/abs/2011A&A...533A..37B}
}

@ARTICLE{lbc,
       author = {{Giallongo}, E. and {Ragazzoni}, R. and {Grazian}, A. and {Baruffolo}, A. and {Beccari}, G. and {de Santis}, C. and {Diolaiti}, E. and {di Paola}, A. and {Farinato}, J. and {Fontana}, A. and {Gallozzi}, S. and {Gasparo}, F. and {Gentile}, G. and {Green}, R. and {Hill}, J. and {Kuhn}, O. and {Pasian}, F. and {Pedichini}, F. and {Radovich}, M. and {Salinari}, P. and {Smareglia}, R. and {Speziali}, R. and {Testa}, V. and {Thompson}, D. and {Vernet}, E. and {Wagner}, R.~M.},
        title = "{The performance of the blue prime focus large binocular camera at the large binocular telescope}",
      journal = {\aap},
         year = 2008,
        month = apr,
       volume = {482},
       number = {1},
        pages = {349-357},
          doi = {10.1051/0004-6361:20078402},
archivePrefix = {arXiv},
       eprint = {0801.1474},
 primaryClass = {astro-ph},
       adsurl = {https://ui.adsabs.harvard.edu/abs/2008A&A...482..349G}
}

@ARTICLE{schlafly2011,
       author = {{Schlafly}, Edward F. and {Finkbeiner}, Douglas P.},
        title = "{Measuring Reddening with Sloan Digital Sky Survey Stellar Spectra and Recalibrating SFD}",
      journal = {\apj},
         year = 2011,
        month = aug,
       volume = {737},
       number = {2},
          eid = {103},
        pages = {103},
          doi = {10.1088/0004-637X/737/2/103},
archivePrefix = {arXiv},
       eprint = {1012.4804},
 primaryClass = {astro-ph.GA},
       adsurl = {https://ui.adsabs.harvard.edu/abs/2011ApJ...737..103S}
}

@ARTICLE{sfd98,
       author = {{Schlegel}, David J. and {Finkbeiner}, Douglas P. and {Davis}, Marc},
        title = "{Maps of Dust Infrared Emission for Use in Estimation of Reddening and Cosmic Microwave Background Radiation Foregrounds}",
      journal = {\apj},
         year = 1998,
        month = jun,
       volume = {500},
       number = {2},
        pages = {525-553},
          doi = {10.1086/305772},
archivePrefix = {arXiv},
       eprint = {astro-ph/9710327},
 primaryClass = {astro-ph},
       adsurl = {https://ui.adsabs.harvard.edu/abs/1998ApJ...500..525S}
}

@ARTICLE{p24,
       author = {{Pace}, Andrew B.},
        title = "{The Local Volume Database: a library of the observed properties of nearby dwarf galaxies and star clusters}",
      journal = {arXiv e-prints},
         year = 2024,
        month = nov,
          eid = {arXiv:2411.07424},
        pages = {arXiv:2411.07424},
          doi = {10.48550/arXiv.2411.07424},
archivePrefix = {arXiv},
       eprint = {2411.07424},
 primaryClass = {astro-ph.GA},
       adsurl = {https://ui.adsabs.harvard.edu/abs/2024arXiv241107424P}
}

@ARTICLE{apt1,
       author = {{Laher}, Russ R. and {Gorjian}, Varoujan and {Rebull}, Luisa M. and {Masci}, Frank J. and {Fowler}, John W. and {Helou}, George and {Kulkarni}, Shrinivas R. and {Law}, Nicholas M.},
        title = "{Aperture Photometry Tool}",
      journal = {\pasp},
         year = 2012,
        month = jul,
       volume = {124},
       number = {917},
        pages = {737},
          doi = {10.1086/666883},
       adsurl = {https://ui.adsabs.harvard.edu/abs/2012PASP..124..737L}
}

@ARTICLE{apt2,
       author = {{Laher}, Russ R. and {Rebull}, Luisa M. and {Gorjian}, Varoujan and {Masci}, Frank J. and {Fowler}, John W. and {Grillmair}, Carl and {Surace}, Jason and {Mattingly}, Sean and {Jackson}, Ed and {Hacopeans}, Eugean and {Hamam}, Nouhad and {Groom}, Steve and {Teplitz}, Harry and {Mi}, Wei and {Helou}, George and {van Eyken}, Julian C. and {Law}, Nicholas M. and {Dekany}, Richard G. and {Rahmer}, Gustavo and {Hale}, David and {Smith}, Roger and {Quimby}, Robert M. and {Ofek}, Eran O. and {Kasliwal}, Mansi M. and {Zolkower}, Jeff and {Velur}, Viswa and {Walters}, Richard and {Henning}, John and {Bui}, Khahn and {McKenna}, Dan and {Kulkarni}, Shrinivas R.},
        title = "{Aperture Photometry Tool Versus SExtractor for Noncrowded Fields}",
      journal = {\pasp},
         year = 2012,
        month = jul,
       volume = {124},
       number = {917},
        pages = {764},
          doi = {10.1086/666507},
       adsurl = {https://ui.adsabs.harvard.edu/abs/2012PASP..124..764L}
}

@ARTICLE{sextractor,
       author = {{Bertin}, E. and {Arnouts}, S.},
        title = "{SExtractor: Software for source extraction.}",
      journal = {\aaps},
         year = 1996,
        month = jun,
       volume = {117},
        pages = {393-404},
          doi = {10.1051/aas:1996164},
       adsurl = {https://ui.adsabs.harvard.edu/abs/1996A&AS..117..393B}
}

@INPROCEEDINGS{Taylor2005,
    author = {{Taylor}, M.~B.},
    title = "{TOPCAT {\amp} STIL: Starlink Table/VOTable Processing Software}",
    booktitle = {Astronomical Data Analysis Software and Systems XIV},
    year = 2005,
    series = {Astronomical Society of the Pacific Conference Series},
    volume = 347,
    editor = {{Shopbell}, P. and {Britton}, M. and {Ebert}, R.},
    month = dec,
    pages = {29},
    adsurl = {http://adsabs.harvard.edu/abs/2005ASPC..347...29T}
}

@ARTICLE{bressan2012,
       author = {{Bressan}, Alessandro and {Marigo}, Paola and {Girardi}, L{\'e}o. and
         {Salasnich}, Bernardo and {Dal Cero}, Claudia and {Rubele}, Stefano and
         {Nanni}, Ambra},
        title = "{PARSEC: stellar tracks and isochrones with the PAdova and TRieste Stellar Evolution Code}",
      journal = {\mnras},
         year = 2012,
        month = nov,
       volume = {427},
       number = {1},
        pages = {127-145},
          doi = {10.1111/j.1365-2966.2012.21948.x},
archivePrefix = {arXiv},
       eprint = {1208.4498},
 primaryClass = {astro-ph.SR},
       adsurl = {https://ui.adsabs.harvard.edu/abs/2012MNRAS.427..127B}
}

@ARTICLE{stet87,
       author = {{Stetson}, Peter B.},
        title = "{DAOPHOT: A Computer Program for Crowded-Field Stellar Photometry}",
      journal = {\pasp},
         year = 1987,
        month = mar,
       volume = {99},
        pages = {191},
          doi = {10.1086/131977},
       adsurl = {https://ui.adsabs.harvard.edu/abs/1987PASP...99..191S}
}

@ARTICLE{allframe,
       author = {{Stetson}, Peter B.},
        title = "{The Center of the Core-Cusp Globular Cluster M15: CFHT and HST Observations, ALLFRAME Reductions}",
      journal = {\pasp},
         year = 1994,
        month = mar,
       volume = {106},
        pages = {250},
          doi = {10.1086/133378},
       adsurl = {https://ui.adsabs.harvard.edu/abs/1994PASP..106..250S}
}



\end{document}